\documentclass[reprint, amsmath, amssymb, aps, pra, superscriptaddress, longbibliography]{revtex4-2}

\usepackage[english]{babel}
\renewcommand{\selectlanguage}[1]{\ignorespaces}
\allowdisplaybreaks

\usepackage{mathtools}
\usepackage{bm}
\usepackage{braket}

\usepackage{graphicx}
\usepackage{dcolumn}
\usepackage[caption=false]{subfig}

\usepackage{tikz}
\usetikzlibrary{arrows.meta, backgrounds}

\usepackage{xcolor}
\usepackage{hyperref}

\hypersetup{
    colorlinks=true,
    linkcolor=blue,
    citecolor=blue,
    urlcolor=blue
}

\definecolor{colorstate}{RGB}{204, 204, 255}
\definecolor{coloroverlap}{RGB}{120, 160, 145}
\definecolor{coloroperator}{RGB}{240, 128, 128}
\definecolor{altcoloroperator}{RGB}{49, 46, 129}
\definecolor{colorvacuum}{RGB}{235, 235, 235}
\definecolor{colorenv}{RGB}{245, 240, 225}
\definecolor{coloraltenv}{RGB}{224, 245, 229}

\tikzset{
    tensor/.style={draw, rectangle, rounded corners=2pt, minimum width=3em, minimum height=1.8em},
    mpstensor/.style={tensor, fill=colorstate!30, semithick},
    mpotensor/.style={tensor, fill=coloroperator!10, semithick},
    altmpotensor/.style={tensor, fill=altcoloroperator!10, semithick},
    envtensor/.style={draw, rectangle, rounded corners=2pt, minimum width=3.8em, minimum height=6.5em, fill=colorenv!40, semithick},
    altenvtensor/.style={draw, rectangle, rounded corners=2pt, minimum width=3.8em, minimum height=6.5em, fill=coloraltenv!40, semithick},
    vactensor/.style={draw=gray!60, rectangle, rounded corners=2pt, fill=gray!10, font=\scriptsize, minimum width=1.8em},
    operator/.append style={rounded corners=3pt}
}

\newcommand{\dd}{\mathrm{d}}
\newcommand{\appref}[1]{\hyperref[#1]{Appendix~\ref*{#1}}}

\def\splitfourb#1,#2,#3,#4\relax{%
  (\hat{b}^{#1})^{\dagger} (\hat{b}^{#2})^{\dagger} \hat{b}^{#3} \hat{b}^{#4}%
}
\newcommand{\fourb}[1]{\splitfourb#1\relax}

\begin{document}

\title{Finite-element matrix product states for continuum models in one dimension}

\author{Akshay Shankar}
\affiliation{Department of Physics and Astronomy, Ghent University, Krijgslaan 281, 9000 Gent,
    Belgium}
\author{Karel Van Acoleyen}
\affiliation{Department of Physics and Astronomy, Ghent University, Krijgslaan 281, 9000 Gent,
    Belgium}

\affiliation{Department~of~Electronics~and~Information~Systems,~Ghent~University,~Technologiepark-Zwijnaarde~126,~9052 Ghent,~Belgium}

\author{Jutho Haegeman}
\affiliation{Department of Physics and Astronomy, Ghent University, Krijgslaan 281, 9000 Gent,
    Belgium}
\date{\today}

\begin{abstract}
    We present a matrix product state framework for simulating one-dimensional quantum many-body systems in the continuum using non-orthogonal single-particle basis sets. By mapping the physical problem to an auxiliary computational space, we show that the resulting many-body overlap operator can be efficiently encoded as a matrix product operator for sufficiently localized orbitals, thereby generalizing a construction that first appeared in \href{https://arxiv.org/abs/2405.10285}{[arXiv:2405.10285]}. This construction recasts the variational ground-state search into a generalized eigenvalue problem, which can be solved using a generalized density matrix renormalization group algorithm. As a primary application, we employ a first-order finite-element expansion to study the ground state properties of the Lieb-Liniger gas in the presence of inhomogeneities. This approach also provides a natural setting for exactly refining the lattice, thereby enabling multigrid optimization strategies for matrix product states.
\end{abstract}

\maketitle

\section{Introduction}

Over the past decades, the development of the density matrix renormalization group (DMRG) algorithm and matrix product states (MPS) has established a computational standard for studying strongly correlated one-dimensional lattice models \cite{schollwockDensitymatrixRenormalizationGroup2011}. However, extending these methods to systems in the continuum, i.e.\ (non-relativistic) field theories such as those describing ultracold atomic gases, remains a significant challenge. Since DMRG is highly optimized for models constructed using orthogonal discrete bases with local interactions, its application to continuum models requires balancing three competing properties: a strict variational principle, basis orthogonality, and spatial locality.

The most ubiquitous approach, finite-difference (FD) discretization, preserves orthogonality and locality by mapping continuous space onto a grid of discrete points \cite{phi4_2004, Stoudenmire2012}. Although this enables the use of standard DMRG, it does not represent a well-defined Hilbert space projection. Consequently, the energy on a discrete grid is not guaranteed to provide an upper bound to that of the continuum ground state, often resulting in non-monotonic convergence. Moreover, in dilute systems, DMRG typically requires excessive computational time to resolve features arising from competing length scales \cite{dolfi_multigrid_2012}.

Alternatively, the continuous matrix product state (cMPS) formalism restores variationality by defining the ansatz directly in the continuum limit \cite{Verstraete2010}. While this has proven successful for studying translationally invariant systems \cite{PhysRevB.90.235142, PhysRevB.91.121108, PhysRevA.96.023609, ganahl2017uniform, PhysRevA.110.023330}, the spatially inhomogeneous case struggles with complications related to the broken gauge redundancy in the parameterization, the computation of the auxiliary environment quantities and the presence of boundaries. This results in a difficult ground state search for which robust and efficient methods remain an ongoing challenge. Hybrid MPS-cMPS algorithms attempt to recycle the benefits of lattice DMRG in combination with multi-grid optimization \cite{ganahl_continuous_2018}, but are sensitive to gauge choices and have not been generalized to open boundary conditions. Meanwhile, direct gradient-based strategies so far lack a suitable choice of preconditioner and therefore have to rely on carefully constructed initial states and a large number of iterations \cite{ganahl2017nonuniform,Tuybens2022,PhysRevB.106.144206}.

To maintain variationality, one might instead project the continuous Hamiltonian onto a subspace spanned by a truncated set of orthogonal basis functions like plane waves \cite{schmoll2023hamiltoniantruncationtensornetworks}. However, their global support destroys the locality of the interactions, leading to unfavorable computational scaling of DMRG \cite{woutersDensityMatrixRenormalization2014}. To resolve this, specialized basis sets like Gausslets \cite{White2017} can be engineered to be simultaneously orthogonal and localized; however, being tailored primarily for quantum chemistry, their application to broader many-body problems remains to be explored. 

On the other hand, finite-element (FE) methods offer a structurally simpler alternative. By expanding the continuous field operators in a basis of localized, piecewise-linear tent functions, the interactions remain local while maintaining a well-defined variational subspace. However, since adjacent tent functions physically overlap, we lose basis orthogonality. Early attempts to combine FE with DMRG relied on explicitly orthogonalizing the basis \cite{White1989}, which destroys the locality. Alternatively, a recent approach directly constructs a basis of model-specific many-body states within the segments \cite{dutta_density_2022}, but requires specialized ``gluing'' conditions that introduce an additional extrapolation parameter.

In this work, we resolve the basis non-orthogonality directly by accounting for it at the many-body level. Instead of transforming the single-particle orbitals, we map the problem to an auxiliary computational space and efficiently encapsulate the basis overlaps within a matrix product operator (MPO). In doing so, the ground-state search is recast as a generalized many-body eigenvalue problem that can be solved using a modified DMRG algorithm \cite{levin_efficient_2026}. While this framework was originally introduced for the study of chiral fermions on a lattice \cite{haegeman_interacting_2024}, we generalize the approach here to encompass bosonic statistics and an arbitrary choice of local orbitals.

The remainder of this paper is organized as follows. Sections \ref{sec:non-orthogonal-basis} and \ref{sec:variational_principle} introduce a general variational framework using a non-orthogonal basis expansion that is applicable to generic bosonic or fermionic Hamiltonians in the continuum. Section \ref{sec:W_construction} details a crucial finding of this work, namely, the exact formulation of the many-body overlap as an efficient MPO. Section \ref{sec:generalized_dmrg} outlines the generalized DMRG algorithm used to variationally optimize the MPS ground state with respect to this overlap metric. Sections \ref{sec:finite_element_mps} and \ref{sec:results} then demonstrate the framework in practice, employing a first-order finite-element basis to capture the continuum ground-state physics of the Lieb-Liniger model in the presence of inhomogeneous potentials. Finally, we reserve our conclusions and discussion for section \ref{sec:outlook}.  The supplementary material can be found in the Appendices \ref{app:hamiltonian_specifics} - \ref{app:other_approaches}.

\section{Non-Orthogonal Basis Expansion}\label{sec:non-orthogonal-basis}

Consider a finite domain, $x \in [-\ell, +\ell]$. We define a set of linearly independent single-particle orbitals, $\{\phi_i(x)\}_{i=1}^L$ that are not necessarily orthogonal, i.e.\ the overlap matrix can be non-trivial
\begin{equation}
    N_{ij} = \int_{-\ell}^{+\ell} \dd x \,\overline{\phi_i(x)} \phi_j(x) \neq \delta_{ij}.
\end{equation}
The matrix $N$ is Hermitian and positive definite, and ---when working with localized orbitals with strictly finite support--- will typically be band diagonal.

We may now define the creation operators
\begin{equation}\label{eq:creation}
    \hat{a}_i^{\dagger} = \int_{-\ell}^{+\ell} \dd x \, \hat{\Psi}^{\dagger}(x) \phi_i(x),
\end{equation}
where $\hat{\Psi}^{\dagger}(x)$ is the fermionic/bosonic creation operator satisfying $[\hat \Psi(x), \hat \Psi^{\dagger}(x')]_{\pm} = \delta(x-x')$. It follows that $\hat{a}_i^{\dagger}\ket{\Omega} = \ket{\phi_i}$ where $\ket{\Omega}$ is the Fock vacuum defined by $\hat{a}_i |\Omega\rangle = \hat{\Psi}(x) \ket{\Omega} = 0$. Due to the non-orthogonality of the basis, the commutation relations are non-canonical:
\begin{equation}\label{eq:commutation_relation}
    [\hat{a}_i, \hat{a}_j^{\dagger}]_{\pm} = N_{ij}.
\end{equation}
With these creation operators, we construct a non-orthogonal Fock basis $\{\ket{n_1,\dots, n_L}\}$,
\begin{equation}
    |n_1, n_2, \dots, n_L\rangle = \left(\prod_{i=1}^L \frac{1}{\sqrt{n_i!}} \, (\hat a^{\dagger}_i)^{n_i} \right) |\Omega \rangle.
\end{equation}
This basis spans a subspace
\begin{displaymath}
    \mathbb{H}^{(L)} = \mathrm{span}\{ |n_1, n_2, \dots, n_L\rangle \mid n_i = 0, 1, \ldots, n_c\} \subseteq \mathbb{H},
\end{displaymath}
of the full many-body Hilbert space $\mathbb{H}$, within which we will construct variational trial states. Here, $n_c=1$ for (spinless) fermions, and can be $+\infty$ for bosons, although it will be truncated to a finite value in practical simulations.

It is now convenient to introduce a set of dual operators \cite{haegeman_interacting_2024,artachoNonorthogonalBasisSets1991}, $\{\hat b^i\}_{i=1}^L$, satisfying
\begin{equation}\label{eq:conjugate_commutation_relation}
    [\hat{b}^i, \hat{a}_j^{\dagger}]_{\pm} = \delta^i_j,
\end{equation}
where
\begin{equation}
    \hat{b}^i = \sum_{j=1}^L (N^{-1})^{ij} \hat{a}_j.
\end{equation}
It then follows that
\begin{equation*}
    \hat{b}^i \ket{n_1,\dots,n_i,\dots,n_L} = \sqrt{n_i} \ket{n_1,\dots,n_i-1,\dots,n_L}.
\end{equation*}
Furthermore, for any state $\ket{\Phi} \in \mathbb{H}^{(L)}$, we have
\begin{equation}\label{eq:b_defn}
    \hat{\Psi}(x) \ket{\Phi} = \sum_{i=1}^L \phi_i(x)\hat{b}^i \ket{\Phi},
\end{equation}
which is a direct consequence of $[\hat{\Psi}(x),\hat{a}_i^\dagger]_{\pm} = \phi_i(x)$.
As a result, in expectation values with respect to states in $\mathbb{H}^{(L)}$, we can effectively project the operators into this subspace. For example, a one-body operator can be re-expressed as
\begin{align}
    \hat{O}^{(1)} & = \int_{-\ell}^{+\ell} \dd x\int_{-\ell}^{+\ell} \dd y \, \hat{\Psi}^\dagger(x) O(x,y) \hat{\Psi}(y)\nonumber                                  \\
                  & \eqsim \int_{-\ell}^{+\ell} \dd x \int_{-\ell}^{+\ell}\dd y \, \overline{\phi_i(x)} O(x,y) \phi_j(y) \; (\hat{b}^i)^\dagger \hat{b}^j\nonumber \\
                  & = \sum_{ij}\langle \phi_i | O |\phi_j \rangle (\hat{b}^i)^\dagger \hat{b}^j.
\end{align}
Generalizing this calculation to two- or higher-body operators, it follows that typical observables projected into $\mathbb{H}^{(L)}$ will appear as normal ordered expressions of $\hat{b}$ and $\hat{b}^{\dagger}$.

\section{Variational Principle as a Generalized Eigenvalue Problem}\label{sec:variational_principle}

Now consider a physical state $\ket{\Phi} \in \mathbb{H}^{(L)}$ that is expanded with respect to the non-orthogonal basis
\begin{equation}
    \ket{\Phi} = \sum_{(n_1,\dots,n_L)} M_{n_1,\dots,n_L} \ket{n_1,\dots,n_L}.
\end{equation}

To perform numerical calculations with $\ket{\Phi}$, it is useful to introduce a computational Fock space $\mathbb{H}^{(L)}_{c}$ constructed from a set of creation operators $\{c_i^\dagger\}$ satisfying the canonical commutation relations (CCR):
\begin{equation}\label{eq:ccr}
    [c_i, c_j^\dagger]_\pm = \delta_{ij}, \qquad [c_i, c_j]_\pm = 0,
\end{equation}
with an associated vacuum state $|\Omega\rangle_{c}^{(L)}$. We denote the corresponding orthogonal Fock basis as
\begin{equation}
    | n_1,  n_2, \dots,  n_L\rangle_{c} = \left(\prod_{i=1}^L \frac{1}{\sqrt{n_i!}} \, (c^{\dagger}_i)^{n_i} \right) |\Omega\rangle_{c}^{(L)}.
\end{equation}
Note that the operators associated with the computational space are denoted without hats to maintain a clear distinction from the physical operators associated with $\mathbb{H}^{(L)}$. The computational space $\mathbb{H}^{(L)}_{c}$ is isomorphic but not isometric or unitarily equivalent to $\mathbb{H}^{(L)}$. In particular, we now introduce a map $W:\mathbb{H}^{(L)}_{c}\to \mathbb{H}^{(L)} \subseteq \mathbb{H}$ that associates each $| n_1,  n_2, \dots,  n_L\rangle_{c}$ with $|n_1,n_2, \ldots, n_L\rangle$. This map is completely specified by
\begin{align}\label{eq:Z_op}
    W c_i^\dagger & = \hat{a}_i^\dagger W, & W |\Omega\rangle_{c}^{(L)} = |\Omega\rangle,
\end{align}
and furthermore satisfies
\begin{align}\label{eq:Z_propertyb}
    \hat{b}^i W = W c_i.
\end{align}
We can now re-express the state as $|\Phi\rangle = W |\Phi\rangle_{c}$ with
\begin{equation}
    |\Phi\rangle_{c} = \sum_{(n_1,\dots,n_L)} M_{n_1,\dots,n_L} |n_1,\dots, n_L\rangle_{c}.
\end{equation}
Correspondingly, the Schrödinger equation $i \partial_t |\Phi\rangle = \hat{H} |\Phi\rangle$ can be expressed in the computational basis. Using Eq.~\eqref{eq:Z_op} and left-multiplying by $W^\dagger$, we obtain:
\begin{equation}
    i W^\dagger W \partial_t |\Phi\rangle_{c} = W^\dagger \hat{H} W |\Phi\rangle_{c} \implies i \mathcal{N} \partial_t |\Phi\rangle_{c} = \mathcal{H} |\Phi\rangle_{c},
    \label{eq:schrodingerprojected}
\end{equation}
where $\mathcal{N} = W^\dagger W$ is the many-body overlap and $\mathcal{H} = W^\dagger \hat{H} W$ is the effective computational Hamiltonian. Here and henceforth, general operators acting on the computation Hilbert space (i.e.\ endomorphisms of $\mathbb{H}_{c}^{(L)}$) will be denoted with a calligraphic font.

In the stationary setting, Eq.~\eqref{eq:schrodingerprojected} reduces to a generalized eigenvalue problem:
\begin{equation}\label{eq:gen_eigval}
    \mathcal{H} |\Phi\rangle_{c} = E {\mathcal{N}} |\Phi\rangle_{c}.
\end{equation}
In order to construct ${\mathcal{N}}$ and ${\mathcal{H}}$ in the computational basis, we now decompose $W = V \mathcal{W}$ where $V:\mathbb{H}^{(L)}_{c} \to \mathbb{H}$ is isometric ($V^\dagger V = \openone$, the identity operator on $\mathbb{H}^{(L)}_{c}$), while $\mathcal{W}:\mathbb{H}^{(L)}_{c} \to\mathbb{H}^{(L)}_{c}$ is invertible (but not unitary) and can be completely expressed in the computational basis. This reduces the many-body overlap to $\mathcal{N} = W^\dagger W = \mathcal{W}^\dagger \mathcal{W}$. Using the property in Eq.~\eqref{eq:Z_propertyb}, it also follows that any Hamiltonian term that is  normal-ordered in $\hat b$ and $\hat b^{\dagger}$ is transformed as
\begin{equation}\label{eq:Z_normal_ordered}
    W^\dagger \big[(\hat{b}^i)^\dagger\big]^m\big[\hat{b}^j\big]^n W = \big[c_i^\dagger\big]^m W^\dagger W \big[c_j\big]^n = \big[c_i^\dagger\big]^m \mathcal{N} \big[c_j]^n.
\end{equation}
It is now apparent that we can mostly forget about $V$ and operate completely in the computational Hilbert space $\mathbb{H}_c^{(L)}$, where a prominent role is played by the many-body overlap operator:
\begin{align*}
    \mathcal{N}_{n_1, n_2, \dots, n_L}^{\, n_1', n_2', \dots, n_L'} & = \langle n_1', \dots, n_L' |n_1, \dots, n_L \rangle                                                    \\
                                                                    & =\ _{c\!}\langle {n}'_1 \dots {n}'_L | \mathcal{W}^\dagger \mathcal{W} | {n}_1 \dots {n}_L \rangle_{c}.
\end{align*}
This fully captures the non-orthogonality of the original modes and serves as the metric of the computational space.

The decomposition $W = V \mathcal{W}$ is not unique, as an arbitrary many-body unitary transformation and its inverse can be absorbed in $V$ and $\mathcal{W}$ respectively. We can reduce this freedom by enforcing that $\mathcal{W}$ acts as a non-orthogonal single-particle basis transformation in $\mathbb{H}_c^{(L)}$. Under this construction, $\mathcal{W}$ leaves the computational vacuum invariant, $\mathcal{W} |\Omega\rangle_c^{(L)} = |\Omega\rangle_c^{(L)}$, and acts linearly on the creation operators (i.e.\ it preserves particle number):
\begin{equation}
    a_i^\dagger = \mathcal{W} {c}_i^\dagger \mathcal{W}^{-1} = \sum_{j} F^j_{\ i} {c}_j^\dagger,\label{eq:defW}
\end{equation}
where we have denoted this specific linear combination as $a_j^\dagger$, which is related but not equal to the physical creation operator $\hat{a}_j^\dagger$ (via $V a_j^\dagger = \hat{a}_j^\dagger V$).
The inner product between the single-particle states $|\phi_i\rangle = \hat{a}_i^\dagger |\Omega \rangle = W {c}_i^\dagger |{\Omega}\rangle_{c}$ then dictates that
\begin{equation}
    \sum_{k} \overline{F^k_{\ i}} F^{k}_{\ j} = (F^{\dagger}F)_{ij} =\langle \phi_i | \phi_j \rangle = N_{ij}.
\end{equation}

While $F$ itself still has unitary freedom at the single-particle level, its structure completely determines whether locality is preserved. For orbitals $\{\phi_i(x)\}$ with sufficiently local support, i.e., $N$ is band diagonal with a bandwidth $R$, a Cholesky decomposition of $N$ yields an upper-triangular $F$ with only $R$ non-zero superdiagonals. Hence, $\mathcal{W}$ acts locally as
\begin{equation}\label{eq:basis_transformation}
    a_i^{\dagger} = \mathcal{W} {c}_i^\dagger \mathcal{W}^{-1} = \sum_{r=0}^R F^{i-r}_{\ \ \ i}  c_{i-r}^{\dagger}\,.
\end{equation}
Crucially, for a small bandwidth $R$, the non-locality of the single-particle orbitals remains modest at the many-body level, namely in such a way that $\mathcal{N}$ admits an efficient MPO representation with a bond dimension that is independent of the total number of basis functions $L$, and only scales with $R$. This will be shown by the explicit construction of an MPO representation of $\mathcal{W}$, and hence $\mathcal{N} = \mathcal{W}^\dagger \mathcal{W}$, in Sec.~\ref{sec:W_construction}.

Following the construction of $\mathcal{N}$, the computational Hamiltonian $\mathcal{H}$ is obtained by first expressing the physical Hamiltonian $H$ in terms of $\hat{b}$ operators, preserving exact locality. As per Eq.~\eqref{eq:Z_normal_ordered}, these operators are then substituted by the canonical $c$ operators, with the overlap $\mathcal{N}$ interleaved between the creation and annihilation components (see \appref{app:hamiltonian_specifics}). The compact MPO structure of $\mathcal{N}$ then ensures that $\mathcal{H}$ maintains an MPO representation with modest bond dimension as well.

Given that both the metric and the Hamiltonian can be efficiently encoded as MPOs, it is natural to restrict the variational class to states $|\Phi \rangle = W |\Phi \rangle_c$ for which the coefficient tensor $M_{n_1,\ldots,n_L}$ is parameterized as an MPS:
\begin{equation*}
    \begin{tikzpicture}[baseline=(current bounding box.center), scale=0.7, semithick,
            tensor/.style={draw, rectangle, rounded corners=2pt, minimum width=1.8em, minimum height=1.4em, fill=colorstate!40, font=\scriptsize}]

        \node[left] at (-1.2, 0) {$M_{n_1, \dots, n_L} = \;$};

        \node[mpstensor] (T1) at (0, 0) {$M^{(1)}$};
        \node[mpstensor] (T2) at (1.6, 0) {$M^{(2)}$};

        \node[mpstensor] (TL) at (5.0, 0) {$M^{(L)}$};

        \draw (T1.east) -- (T2.west);

        \draw (T2.east) -- (TL.west) node[midway, fill=white, inner sep=2pt, font=\small] {$\dots$};

        \draw (T1.south) -- ++(0,-0.5) node[below, font=\small] {$n_1$};
        \draw (T2.south) -- ++(0,-0.5) node[below, font=\small] {$n_2$};
        \draw (TL.south) -- ++(0,-0.5) node[below, font=\small] {$n_L$};
    \end{tikzpicture},
\end{equation*}
where $M^{(i)}$ denotes the local tensor at site $i$.

This formulation allows the ground-state search to be recast as a generalized many-body eigenvalue problem that can be efficiently solved using a modified DMRG algorithm, as shown in Sec.~\ref{sec:generalized_dmrg}.

\section{Constructing the many-body overlap MPO}\label{sec:W_construction}
We now show that, for a localized single-particle basis with band diagonal overlap matrix $N$, the many-body overlap operator admits a compact MPO representation:
\begin{equation}\label{eq:overlap_mpo}
    \mathcal{N}_{n_1, \dots, n_L}^{n_1', \dots, n_L'} = \prod_{i=1}^L \mathcal{N}^{(i) n_i'}_{\, \, n_i},
\end{equation}
where $\mathcal{N}^{(i) n_i'}_{n_i}$ is a matrix at each site with some dimension $\chi_{i-1} \times \chi_i$ that does not scale with $L$.

In order to construct $\mathcal{N}$, we first construct the matrix elements of $\mathcal{W}$. To this end, we use $\mathcal{W} c_i^\dagger = a_i^\dagger \mathcal{W}$ according to Eq.~\eqref{eq:defW} in order to write
\begin{align}
    \mathcal{W}|\{n_i\}\rangle_{c}
     & = \prod_{i=1}^L \frac{1}{\sqrt{n_i!}} \mathcal{W} ( c_i^\dagger)^{n_i} |\Omega\rangle_{c}^{(L)} \nonumber                                                      \\
     & = \prod_{i=1}^L \frac{1}{\sqrt{n_i!}} ( a_i^\dagger)^{n_i} |\Omega\rangle_{c}^{(L)} \nonumber                                                                  \\
     & = \prod_{i=1}^L \left[ \frac{1}{\sqrt{n_i!}} \left( \sum_{r=0}^R F^{i-r}_{\ \ \ i}  c_{i-r}^{\dagger} \right)^{n_i} \right] |\Omega\rangle_{c}^{(L)} \nonumber \\
     & = \prod_{i=1}^L \Bigg[ \sum_{\{k_{i,r}\}} \delta_{n_i, \sum_r k_{i,r}} \frac{1}{\sqrt{n_i!}} \nonumber                                                         \\
     & \qquad \times \binom{n_i}{{\{k_{i,r}\}}} \prod_{r=0}^R (F^{i-r}_{\ \ \ i}  c_{i-r}^{\dagger})^{k_{i,r}} \Bigg] |\Omega\rangle_{c}^{(L)},
\end{align}
where we have performed a multinomial expansion by introducing summation variables $k_{i,r}$ subject to the particle number constraint $\sum_{r=0}^R k_{i,r} = n_i$, as enforced by the Kronecker delta. By introducing this factor explicitly, the summation range of the different $k_{i,r}$ can be chosen from $0$ to $n_c$ independent of $i$.

This enables us to reorganize the terms and collect all creation operators $c_{i}^\dagger$ on a site $i$, counted by a new index, $n_i'$. We thus obtain
\begin{align}\label{eq:multinomial_ks}
    \mathcal{W}|\{n_i\}\rangle_{c} & = \sum_{\{k_{i,r}\}} \xi(\{k_{i,r}\}) \prod_{i=1}^L \bigg[ \delta_{n_i, \sum_r k_{i,r}} \sqrt{\frac{n_i'!}{n_i!}} \nonumber \\
                                   & \quad \times \binom{n_i}{\{k_{i,r}\}} \prod_{r=0}^R (F^{i-r}_{\ \ \ i})^{k_{i,r}} \bigg] |\{ n_i'\}\rangle_{c},
\end{align}
with $n_i' = \sum_{r=0}^R k_{i+r,r}$. Note that, in the case of bosons, if we introduce a cutoff $n_i \leq n_c$,  a target site $i$ can accumulate creation operators from up to $R$ neighboring sites to its right. As a result, an exact representation of $\mathcal{W}$ ---required to correctly compute $\mathcal{N}=\mathcal{W}^\dagger \mathcal{W}$--- necessitates keeping $n_i' \leq (R+1) n_c$ modes on each site in the output of $\mathcal{W}$. For fermions on the other hand, where only $n_i, n_i' = 0, 1$ are allowed, the reorganization introduces an extra sign factor $\xi(\{k_{i,r}\})$ that is yet to be determined.

\begin{figure}[t]
    \centering
    \resizebox{\columnwidth}{!}{
        \begin{tikzpicture}[
                qbox/.style={draw, dashed, draw opacity=0.5, fill=brown!20, fill opacity=0.6, rounded corners=2pt},
                xscale=2.2,
                yscale=1.1,
                >=Stealth,
                site line/.style={ultra thick},
                boundary line/.style={gray!50, dashed, thick},
                flow box/.style={draw=none, fill=gray!20, fill opacity=0.8, text opacity=1, rounded corners=3pt, inner sep=3pt, font=\scriptsize},
                k label/.style={font=\small, anchor=west, xshift=2pt, inner sep=1pt}
            ]

            \def\yoffset{-0.3}

            \foreach \x in {-0.5, 0.5, 1.5, 2.5} {
                    \draw[boundary line] (\x, -0.1) -- (\x, 6.5);
                }

            \foreach \x/\label in {-1/i-2, 0/i-1, 1/i, 2/i+1, 3/i+2} {
                    \draw[site line] (\x, 0) -- (\x, 0.4);
                    \node[below=4pt, font=\small] at (\x, 0) {$\label$};
                }

            flows ---
            \begin{scope}[yshift=\yoffset cm]
                \draw[->, thick] (2.8, 6.5) node[k label] {$k_{i+2, 2}$} -- (1, 6.5);
                \draw[->, thick] (1.8, 5.7) node[k label] {$k_{i+1, 1}$} -- (1, 5.7);
                \draw[->, thick] (1.8, 4.5) node[k label] {$k_{i+1, 2}$} -- (0, 4.5);
                \draw[->, thick] (0.8, 3.7) node[k label] {$k_{i, 1}$} -- (0, 3.7);
                \draw[->, thick] (0.8, 2.5) node[k label] {$k_{i, 2}$} -- (-1, 2.5);
                \draw[->, thick] (-0.25, 1.7) node[k label] {$k_{i-1, 1}$} -- (-1, 1.7);

                \begin{scope}[on background layer]
                    \fill[qbox] (1.25, 5.2) rectangle (1.75, 7.0);
                    \node[anchor=south, font=\small] at (1.5, 5.85) {$Q_{i, 1}$};

                    \fill[qbox] (1.25, 3.2) rectangle (1.75, 5.0);
                    \node[anchor=south, font=\small] at (1.5, 3.85) {$Q_{i, 2}$};

                    \fill[qbox] (0.25, 3.2) rectangle (0.75, 5.0);
                    \node[anchor=south, font=\small] at (0.5, 3.85) {$Q_{i-1, 1}$};

                    \fill[qbox] (0.25, 1.2) rectangle (0.75, 3.0);
                    \node[anchor=south, font=\small] at (0.5, 1.85) {$Q_{i-1, 2}$};
                \end{scope}
            \end{scope}
        \end{tikzpicture}
    }
    \caption{Visual schematic of the particle-flow picture with $R=2$. The indices $k_{i,r}$ count the particles originating at site $i$ and travel a distance $r$, while the virtual indices $Q_{i,r}$ count the total particle flow across the boundary between sites $i$ and $i+1$ that travel to $i-r+1$.}
    \label{fig:particle_flow}
\end{figure}
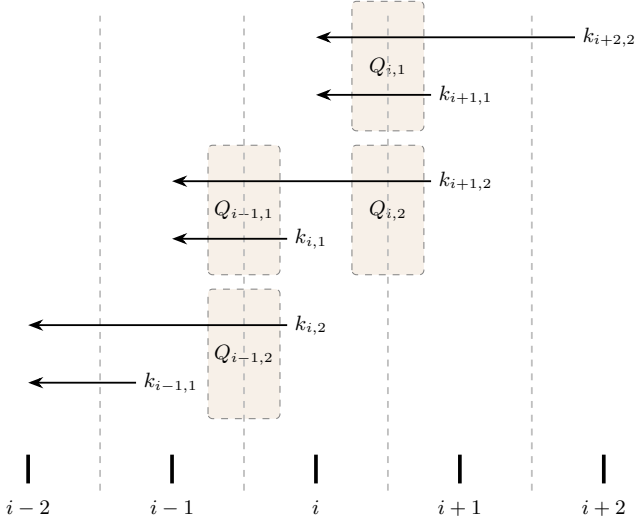

At this point, it is useful to view the algebraic re-arrangement of the creation operators in Eq.~\eqref{eq:multinomial_ks} in terms of virtual processes that displace particles from their starting sites in order to realize the basis transformation. In this picture, the index $k_{i,r}$ counts the number of particles that originate at site $i$ and end up at site $i-r$. However, in order to construct an MPO representation, we require a description that does not directly depend on long-range information, but rather on variables that are shared across neighboring sites. These shared variables will then take the role of the virtual indices of the MPO tensors. To this end, we define $\vec{Q}_i = (Q_{i,1}, \dots, Q_{i,R})$ on the link between sites $i$ and $i+1$, where the value $ Q_{i,r}$ represents the total number of particles crossing the link that originated at some $j>i$ and will end up at site $i - r + 1$, as visualized in Fig.~\ref{fig:particle_flow}. This only tracks how many particles pass through a link regardless of exactly where they originated. From this definition, it follows that:
\begin{subequations}\label{eq:transition_rules}
    \begin{align}
        n_i'       & = k_{i,0} + Q_{i,1} \label{eq:transition_rule_1}                                     \\
                   & \text{\scriptsize (Particles ending up at site $i$)} \nonumber                       \\[1ex]
        Q_{i-1, r} & = k_{i,r} + Q_{i, r+1} \label{eq:transition_rule_2}                                  \\
                   & \text{\scriptsize (Particles starting from $j \geq i$ ending up at $i-r$)} \nonumber
    \end{align}
\end{subequations}
for $1 \leq r \leq R$, with $Q_{i, R+1} \equiv 0$ and boundary conditions $\vec{Q}_0 = \vec{Q}_L = \vec{0}$. These relations allow the original variables $\{k_{i,r}\}$ to be recovered from the physical output $n_i'$ and the virtual indices $\vec{Q}_{i-1}$ and $\vec{Q}_{i}$. Specifically, by also defining $Q_{i-1,0} \equiv n_i'$, we have $k_{i,r} = Q_{i-1,r} - Q_{i,r+1}$ for all $r=0,\ldots, R$. Summing all equations in Eq.~\eqref{eq:transition_rules} yields particle number conservation
\begin{equation}
    \sum_{r=1}^{R} Q_{i-1,r} + n_i' = \sum_{r} k_{i,r} + \sum_{r=1}^{R} Q_{i,r} = n_i + \sum_{r=1}^{R} Q_{i,r}.
\end{equation}

Additionally, this formulation enables a factorization of the fermionic sign factor $\xi(\{k_{i,r}\}) \equiv \xi(\{Q_{i,r}\}) = \prod_i \xi_i$ by tracking the number of particle trajectories that cross each other. To count these crossings locally at site $i$, we consider the trajectory of a particle originating at $i$ against those originating further to the right ($j>i$). A crossing between these occurs only if the latter end up strictly to the left of the former. Thus, particles from $i$ that end up at $i-r$ (as counted by $k_{i,r}$) cross with particles from the right ending up at $i - r' + 1$ (as counted by $Q_{i,r'}$) provided that $i - r' + 1 < i - r$, which simplifies to $r' \ge r + 2$. The local sign factor is then given by:
\begin{equation}\label{eq:local_sign_factor}
    \xi_i = \begin{cases}
        +1                                                                   & \text{for bosons,}   \\
        (-1)^{\sum_{r=0}^{R} k_{i,r} \left(\sum_{r'=r+2}^R Q_{i, r'}\right)} & \text{for fermions.}
    \end{cases}
\end{equation}
Putting everything together, we obtain
\begin{equation}
    \mathcal{W} |\{n_i\}\rangle_{c} = \sum_{\{n_i'\}} \sum_{\{\vec{Q}_i\}} \prod_{i=1}^L \mathcal{W}^{(i)}_{n_i', \, n_i, \, \vec{Q}_{i-1}, \, \vec{Q}_i} | \{n_i'\} \rangle_c,
\end{equation}
which is exactly an MPO representation, with local tensors given as follows, using $k_{i,r} = Q_{i-1,r} - Q_{i,r+1}$:
\begin{equation}
    \begin{aligned}
         & \mathcal{W}^{(i)}_{n_i', n_i, \vec{Q}_{i-1}, \vec{Q}_{i}} =                                                                                                                                                                                                                                                             \\
         & \ \begin{cases}
                 \xi_i \sqrt{\frac{n_i'!}{n_i!}} \binom{n_i}{k_{i,0}, \dots, k_{i,R}} \prod_{r=0}^R (F^{i-r}_{\ \ \ i})^{k_{i,r}} & \text{if } \substack{\sum_r k_{i,r} = n_i \\ \forall r, \, k_{i,r} \geq 0} \\
                 0                                                                                                                & \text{otherwise.}
             \end{cases}
    \end{aligned}
\end{equation}

It is now apparent from the definitions in Eq.~\eqref{eq:transition_rules}, that the virtual indices are bounded as $ 0\leq Q_{i,r} \le n_c (R-r+1)$, resulting in a maximal bond dimension $\chi = \prod_{r=1}^R (n_c r + 1)$. It follows that the many-body overlap $\mathcal{N} = \mathcal{W}^{\dagger}\mathcal{W}$ has a maximal bond dimension $\chi = (\prod_{r=1}^R (n_c r + 1))^2$. Thus, we have identified an MPO representation of $\mathcal{N}$ that does not scale with the number of basis functions, $L$ and retains a low bond dimension for a sufficiently local set of single-particle orbitals $\{\phi_i(x)\}$ (i.e.\ with a sufficiently small bandwidth $R$).

\section{Generalized Density Matrix Renormalization Group}\label{sec:generalized_dmrg}

With both the computational Hamiltonian ${\mathcal{H}}$ and the many-body overlap ${\mathcal{N}}$ represented as MPOs, the ground state may be determined by solving the many-body generalized eigenvalue problem established in Eq.~\eqref{eq:gen_eigval} using standard tensor network algorithms.

We approach this using a generalization of the DMRG algorithm \cite{levin_efficient_2026}, which sequentially optimizes each local tensor $M^{(i)}$ keeping all other tensors fixed. For any MPO $\hat{O} = \prod_i O^{(i)}$, the global expectation value is reduced to a local quadratic form, $_{\,c\!}\langle\Phi|\hat{O}|\Phi\rangle_c = {M^{(i)}}^\dagger \hat{O}_{\text{eff}}^{(i)} M^{(i)}$ at site $i$, where the effective operator is defined as shown in Eq.~\eqref{eq:eff_operator_defn}.
\begin{subequations}
    \label{eq:eff_operator_defn}
    \begin{equation}
        \label{eq:eff_operator_generic}
        O_{\text{eff}}^{(i)} =
        \begin{tikzpicture}[baseline=(current bounding box.center), scale=0.6, semithick]
            \node[envtensor] (L) at (0,0.7) {$O_L^{(i-1)}$};
            \node[mpotensor] (W) at (2.5,0.7) {$O^{(i)}$};
            \node[envtensor] (R) at (5,0.7) {$O_R^{(i+1)}$};
            \draw (L.east |- 0,2.2) -- ++(0.6,0);
            \draw (R.west |- 0,2.2) -- ++(-0.6,0);
            \draw (L.east |- W.center) -- (W.west);
            \draw (W.east) -- (R.west |- W.center);
            \draw (L.east |- 0,-0.8) -- ++(0.6,0);
            \draw (R.west |- 0,-0.8) -- ++(-0.6,0);
            \draw (W.north) -- ++(0,0.7);
            \draw (W.south) -- ++(0,-0.7);
        \end{tikzpicture}
    \end{equation}
    \begin{equation}
        \label{eq:left_env_update}
        \begin{tikzpicture}[baseline=(current bounding box.center), scale=0.55, semithick]
            \node[envtensor] (L) at (0,0.7) {$O_L^{(i)}$};
            \foreach \y in {2.2, 0.7, -0.8} { \draw (L.east |- 0,\y) -- ++(0.7,0); }
        \end{tikzpicture}
        =
        \begin{tikzpicture}[baseline=(current bounding box.center), scale=0.55, semithick]
            \node[envtensor] (Lprev) at (0,0.7) {$O_L^{(i-1)}$};
            \node[mpstensor] (A) at (2.7, 2.2) {$M^{(i)}$};
            \node[mpotensor] (W) at (2.7, 0.7) {$O^{(i)}$};
            \node[mpstensor] (Ab) at (2.7, -0.8) {$\bar{M}^{(i)}$};
            \draw (Lprev.east |- A) -- (A.west);
            \draw (Lprev.east |- W) -- (W.west);
            \draw (Lprev.east |- Ab) -- (Ab.west);
            \draw (A.east) -- ++(0.7,0);
            \draw (W.east) -- ++(0.7,0);
            \draw (Ab.east) -- ++(0.7,0);
            \draw (A) -- (W) -- (Ab);
        \end{tikzpicture}
    \end{equation}
    \begin{equation}
        \label{eq:right_env_update}
        \begin{tikzpicture}[baseline=(current bounding box.center), scale=0.55, semithick]
            \node[envtensor] (R) at (0,0.7) {$O_R^{(i)}$};
            \foreach \y in {2.2, 0.7, -0.8} { \draw (R.west |- 0,\y) -- ++(-0.7,0); }
        \end{tikzpicture}
        =
        \begin{tikzpicture}[baseline=(current bounding box.center), scale=0.55, semithick]
            \node[envtensor] (Rnext) at (2.7,0.7) {$O_R^{(i+1)}$};
            \node[mpstensor] (A) at (0, 2.2) {$M^{(i)}$};
            \node[mpotensor] (W) at (0, 0.7) {$O^{(i)}$};
            \node[mpstensor] (Ab) at (0, -0.8) {$\bar{M}^{(i)}$};
            \draw (Rnext.west |- A) -- (A.east);
            \draw (Rnext.west |- W) -- (W.east);
            \draw (Rnext.west |- Ab) -- (Ab.east);
            \draw (A.west) -- ++(-0.7,0);
            \draw (W.west) -- ++(-0.7,0);
            \draw (Ab.west) -- ++(-0.7,0);
            \draw (A) -- (W) -- (Ab);
        \end{tikzpicture}
    \end{equation}
\end{subequations}
Consequently, the optimization at site $i$ reduces to solving the simpler generalized eigenvalue problem:
\begin{equation}\label{eq:gen_eigval_local}
    {\mathcal{H}}_{\text{eff}}^{(i)} M^{(i)} = E {\mathcal{N}}_{\text{eff}}^{(i)} M^{(i)}.
\end{equation}

To maintain favorable computational scaling, we avoid explicitly inverting the effective metric $\mathcal{N}_{\text{eff}}^{(i)}$. Instead, we use the Locally Optimal Block Preconditioned Conjugate Gradient (LOBPCG) algorithm \cite{duerschRobustEfficientImplementation2018}, a matrix-free approach that solves the generalized problem without the overhead of inversion.

The computational complexity of such a DMRG scheme scales as $\mathcal{O}(L d \chi D^3)$ \cite{schollwockDensitymatrixRenormalizationGroup2011}, where $L$ is the number of basis functions, $d$ is the dimension of the physical space of the MPS, $\chi$ is the maximal MPO bond dimension of $\mathcal{H}$ and $\mathcal{N}$, and $D$ is the MPS bond dimension. For single-particle orbitals with a bandwidth $R$ and a particle cutoff, $n_c$, we have $d=n_c+1$ and $\chi \sim R!^2 \, d^{2R}$ resulting in a scaling $\mathcal{O}(R!^2\, d^{2R+1} L D^3)$.

\subsection{Extensivity of the norm}
A practical challenge in implementing this scheme arises from the fact that the many-body overlap has an expectation value ${}_{\,c\!}\langle \Phi | \mathcal{N} | \Phi \rangle_c$ that is extensive in $L$ when we work with the standard canonical forms \cite{schollwockDensitymatrixRenormalizationGroup2011} for the MPS tensors.

As a result, the elements of the left and right environments $\mathcal{N}_{L}^{(i)}$ and $\mathcal{N}_R^{(i)}$ defined in Eq.~\eqref{eq:left_env_update} and Eq.~\eqref{eq:right_env_update} tend to grow or shrink exponentially across the chain. This scaling can lead to numerical instability due to floating-point underflow or overflow when formulating the local generalized eigenvalue problem in Eq.~\eqref{eq:gen_eigval_local}. We address this by noting that the effective Hamiltonian ${\mathcal{H}}$ incorporates the same overlap contributions as the metric ${\mathcal{N}}$ as shown in Eq.~\eqref{eq:Z_normal_ordered}. Consequently, the exponentially large scaling factors are common to both ${\mathcal{H}}_{\text{eff}}^{(i)}$ and ${\mathcal{N}}_{\text{eff}}^{(i)}$, rendering the local energy intensive and well-conditioned.

In practice, we thus maintain numerical stability by rescaling the environment tensors at each site during the DMRG sweep such that their entries remain $\mathcal{O}(1)$. The logarithm of the extensive scaling factors are maintained separately and are only used when computing the energy. This approach allows for robust optimization without requiring a non-trivial canonicalization of the MPS tensors with respect to the metric ${\mathcal{N}}$.

\subsection{Error measure}

To monitor the proximity of the parameterized state $|\Phi(\{M^{(i)}\})\rangle$ to a variational stationary point, we utilize the norm of the energy gradient with respect to the local tensors $\{M^{(i)}\}$. The energy functional is defined as:
\begin{equation}
    E(\{M^{(i)}\}) = \frac{\langle \Phi(\{M^{(i)}\}) | {\mathcal{H}} | \Phi(\{M^{(i)}\}) \rangle}{\langle \Phi(\{M^{(i)}\}) | {\mathcal{N}} | \Phi(\{M^{(i)}\}) \rangle}.
\end{equation}
Taking the derivative with respect to the conjugate local tensor $\bar{M}^{(i)}$ yields the local gradient:
\begin{align}
    |\nabla_i E\rangle & = \frac{1}{\langle \Phi | {\mathcal{N}} | \Phi \rangle} \left( \frac{\partial \langle \psi | {\mathcal{H}} | \Phi \rangle}{\partial \bar{M}^{(i)}} - E \frac{\partial \langle \Phi | {\mathcal{N}} | \Phi \rangle}{\partial \bar{M}^{(i)}} \right) \nonumber \\
                       & = \frac{({\mathcal{H}}_{\text{eff}}^{(i)} - E {\mathcal{N}}_{\text{eff}}^{(i)}) M^{(i)}}{\langle \Phi | {\mathcal{N}} | \Phi \rangle}.
\end{align}
The gradient norm vanishes exactly when the stationary condition Eq.~\eqref{eq:gen_eigval_local} is satisfied, providing a more robust convergence metric than the change in energy between subsequent DMRG iterations.

\section{Finite element Matrix product states}\label{sec:finite_element_mps}
In the rest of this work, we specifically consider the Lieb-Liniger Hamiltonian \cite{Lieb1963} in the presence of an external potential $V(x)$,
\begin{align}\label{eq:LiebLiniger}
    \hat{H} & = \int_{-\ell}^{+\ell} \dd x \, \hat{\Psi}^{\dagger}(x) \left[ -\frac{1}{2} \partial_x^2 + V(x) - \mu \right] \hat{\Psi}(x) \nonumber \\
            & \quad + g\int_{-\ell}^{+\ell} \dd x \, \hat{\Psi}^{\dagger}(x)\hat{\Psi}^{\dagger}(x)\hat{\Psi}(x)\hat{\Psi}(x),
\end{align}
where we set $\hbar = m = 1$ and the field operators satisfy the bosonic CCR $[\hat{\Psi}(x), \hat{\Psi}^{\dagger}(x')] = \delta(x - x')$.

\subsection{First order linear elements}
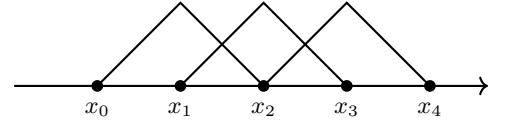
\begin{figure}[!ht]
    \centering
    \begin{tikzpicture}[scale=1.1, thick]
        \draw[->] (-0.,0) -- (5.7,0);

        \draw[black] (1,0) -- (2,1) -- (3,0);
        \draw[black] (2,0) -- (3,1) -- (4,0);
        \draw[black] (3,0) -- (4,1) -- (5,0);

        \foreach \x in {0,1,2,3,4} {
                \fill (\x+1, 0) circle (2pt);
                \node[below=3pt] at (\x+1, 0) {\small $x_{\x}$};
            }
    \end{tikzpicture}
    \caption{First order finite element basis for $L=3$. \label{fig:tent}}
\end{figure}

Given a finite domain, $x \in [-\ell, +\ell]$, we define a set of real-valued piecewise-linear tent functions $\{\phi_i(x)\}_{i=1}^L$ centered over points on a uniform grid with spacing $\Delta x = 2 \ell/(L+1)$ so that $x_i = -\ell + i\Delta x $, i.e.\
\begin{equation}\label{eq:tent_fn}
    \phi_i(x) =
    \begin{cases}
        (x - x_{i-1})/h, & x \in [x_{i-1}, x_i]   \\
        (x_{i+1} - x)/h, & x \in (x_{i}, x_{i+1}] \\
        0,               & \text{otherwise}
    \end{cases}
\end{equation}
where $h = \sqrt{2(\Delta x)^3/3}$ is a normalization factor. These functions are depicted in Fig.~\ref{fig:tent}.

The resulting single-particle overlap matrix $N_{ij} = \langle \phi_i | \phi_j \rangle$ is symmetric and tridiagonal:
\begin{equation}
    N_{ij} = \int_{-\ell}^{+\ell} \phi_i(x) \phi_j(x)\,\mathrm{d}x = \begin{cases}
        1,   & i = j             \\
        1/4, & |i - j| = 1       \\
        0,   & \text{otherwise.}
    \end{cases}
\end{equation}
Since $N$ has a small bandwidth $R = 1$, the MPO representation of $\mathcal{N}$ simplifies significantly. The vector of virtual indices $\vec{Q}_i$ reduces to a single scalar $Q_i$ that counts the number of particles that originate at site $i+1$ and end up at site $i$. Consequently, the conservation of particles simplifies to $n_i' = n_i - Q_{i-1} + Q_i$, with the boundary conditions $Q_0 = Q_L = 0$. The local MPO tensor $\mathcal{W}^{(i)}$ can then be expressed as:
\begin{displaymath}
    \begin{aligned}
        \begin{tikzpicture}[baseline=(current bounding box.center), scale=0.45, semithick]

            \node[mpotensor] (T) at (0,0) {\small $\mathcal{W}^{(i)}$};
            \draw (T.north) -- ++(0,0.7) node[above] {\scriptsize $n_i$};
            \draw (T.west) -- ++(-0.7,0) node[left] {\scriptsize $Q_{i-1}$};
            \draw (T.south) -- ++(0,-0.7) node[below] {\scriptsize $n_i'$};
            \draw (T.east) -- ++(0.7,0) node[right] {\scriptsize $Q_i$};

        \end{tikzpicture}
         & = \sqrt{\frac{n_i'!}{n_i!}} \binom{n_i}{Q_{i-1}} F_{i,i}^{n_i - Q_{i-1}} \\[-3ex]
         & \times F_{i-1, i}^{Q_{i-1}} \; \delta_{n_i', n_i - Q_{i-1} + Q_i}
    \end{aligned}
\end{displaymath}
where $0 \leq Q_i \leq n_c$, such that $n_i' = n_i + Q_i - Q_{i-1} \leq 2n_c$. With physical legs $n_i, \tilde{n}_i$ and virtual bonds $\mathbb{Q} = (Q, \tilde{Q})$, it follows that $\mathcal{N} = \mathcal{W}^{\dagger}\mathcal{W}$ has local MPO tensors
\begin{displaymath}
    \begin{aligned}
        \begin{tikzpicture}[baseline=(current bounding box.center), scale=0.45, semithick]

            \node[mpotensor] (T) at (0,0) {\small $\mathcal{N}^{(i)}$};

            \draw (T.north) -- ++(0,0.7) node[above] {\scriptsize $n_i$};
            \draw (T.west) -- ++(-0.7,0) node[left] {\scriptsize $\mathbb{Q}_{i-1}$};
            \draw (T.south) -- ++(0,-0.7) node[below] {\scriptsize $\tilde n_i$};
            \draw (T.east) -- ++(0.7,0) node[right] {\scriptsize $\mathbb{Q}_i$};
        \end{tikzpicture}
         & = \frac{(n_i - Q_{i-1} + Q_i)!}{\sqrt{n_i! \tilde{n}_i!}}                               \\[-4ex]
         & \times \binom{n_i}{Q_{i-1}} \binom{\tilde{n}_i}{\tilde{Q}_{i-1}}                        \\
         & \times F_{i,i}^{n_i - Q_{i-1}} (F_{i,i}^*)^{\tilde{n}_i - \tilde{Q}_{i-1}}              \\
         & \times F_{i-1, i}^{Q_{i-1}} (F_{i-1, i}^*)^{\tilde{Q}_{i-1}}                            \\
         & \times \delta_{n_i - \tilde{n}_i, \, (Q_{i-1} - \tilde{Q}_{i-1}) - (Q_i - \tilde{Q}_i)}
    \end{aligned}
\end{displaymath}
with bond dimension $\chi = (n_c + 1)^2$.

We can now construct the Hamiltonian by first substituting Eq.~\eqref{eq:b_defn} in Eq.~\eqref{eq:LiebLiniger} to get:
\begin{align}
    \hat{H} = \sum_{ij} t_{ij} (\hat{b}^i)^{\dagger} \hat{b}^j + \sum_{ijkl}  U_{ijkl} \fourb{i,j,k,l},
\end{align}
where $t = t^k + t^v - \mu t^c$ such that
\begin{subequations}
    \begin{align}
        t^k_{ij} & = \frac{1}{2}\int_{-\ell}^{+\ell} \dd x \, \partial_x\phi_i(x) \partial_x\phi_j(x) \\
        t^v_{ij} & = \int_{-\ell}^{+\ell} \dd x \, V(x)\phi_i(x)\phi_j(x)                             \\
        t^c_{ij} & = \int_{-\ell}^{+\ell} \dd x \, \phi_i(x)\phi_j(x)                                 \\
        U_{ijkl} & = g\int_{-\ell}^{+\ell} \dd x \, \phi_i(x) \phi_j(x) \phi_k(x) \phi_l(x).
    \end{align}
\end{subequations}
It follows from Eq.~\eqref{eq:Z_normal_ordered} that the effective Hamiltonian in the computational basis is obtained as
\begin{align}\label{eq:LiebLiniger_comp}
    \mathcal{H} = \sum_{ij} t_{ij} \, c_i^{\dagger} \mathcal{N}c_j +  \sum_{ijkl}  U_{ijkl} \, c_i^{\dagger} c_j^{\dagger}\mathcal{N} c_k c_l.
\end{align}
Since $\phi_i(x)$ only has local support on the interval $[x_{i-1}, x_{i+1}]$, all the matrix elements mentioned above vanish unless the site indices are identical ($i=j$) or adjacent ($i=j\pm 1$). This allows us to write down a compact MPO representation of the Hamiltonian as shown in
\appref{app:hamiltonian_specifics}.

\subsection{Multigrid optimization}
\label{sec:refinement}
It is known that the optimization of continuum models using DMRG often encounters convergence difficulties when the system is discretized onto a lattice due to the presence of competing length scales \cite{dolfi_multigrid_2012}. In such scenarios, a multigrid approach is found to be highly effective \cite{dolfi_multigrid_2012, ganahl_continuous_2018}: coarse details are resolved on a sparse grid and the solution is used as an initial state for optimization on a refined grid. It turns out that the tent functions possess an interesting property that makes them particularly compelling for this purpose.
\subsubsection{Refinement Property}

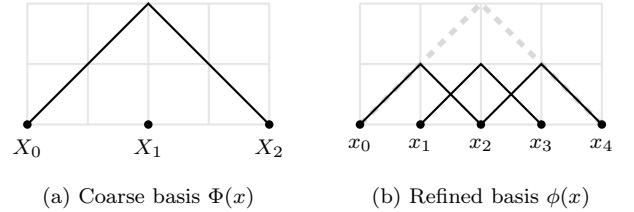
\begin{figure}[ht]
    \centering
    \begin{tikzpicture}[scale=0.8, thick]
        \begin{scope}[xshift=0cm]
            \draw[gray!20] (0,0) grid (4,2);
            \draw[black] (0,0) -- (2,2) -- (4,0);

            \foreach \x in {0,1,2} {
                    \fill (2*\x,0) circle (2pt);
                    \node[below=2pt] at (2*\x,0) {\footnotesize $X_{\x}$};
                }
            \node[below=20pt] at (2,0) {\footnotesize (a) Coarse basis $\Phi(x)$};
        \end{scope}

        \begin{scope}[xshift=5.5cm]
            \draw[gray!20] (0,0) grid (4,2);

            \draw[gray, line width=2pt, opacity=0.3, dashed] (0,0) -- (2,2) -- (4,0);

            \draw[black] (0,0) -- (1,1) -- (2,0);
            \draw[black] (1,0) -- (2,1) -- (3,0);
            \draw[black] (2,0) -- (3,1) -- (4,0);

            \foreach \x in {0,1,2,3,4} {
                    \fill (\x,0) circle (2pt);
                    \node[below=2pt] at (\x,0) {\footnotesize $x_{\x}$};
                    3            }
            \node[below=20pt] at (2,0) {\footnotesize (b) Refined basis $\phi(x)$};
        \end{scope}
    \end{tikzpicture}
    \caption{Refinement of the tent function basis\label{fig:refinement}}
\end{figure}

Consider a set of tent functions $\{\Phi_i(x)\}_{i=1}^L$ centered at equally spaced points $\{X_i\}_{i=1}^L$, where the domain is fixed by the boundary points $X_0=-\ell$ and $X_{L+1}=+\ell$. We can now construct a finer grid $\{x_i\}_{i=1}^{2L+1}$ in the interior of the domain by introducing a new vertex at the midpoint of every coarse interval, yielding the correspondence $X_i = x_{2i}$ for $i=1, \ldots, L$. Associating the tent functions $\{\phi_i(x)\}_{i=1}^{2L+1}$ with this new grid, as depicted in Fig.~\ref{fig:refinement}, the coarse basis functions can be expanded as:
\begin{equation}\label{eq:tent_multigrid}
    \Phi_i(x) = \frac{1}{2\sqrt{2}}(\phi_{2i-1}(x) + 2\phi_{2i}(x) + \phi_{2i+1}(x)).
\end{equation}
This property ensures that any many-body state represented on the coarse grid can be mapped \textit{exactly} onto the fine grid, i.e.\ $\mathbb{H}^{(L)} \subset \mathbb{H}^{(2L+1)}$, a feature not shared by finite-difference discretizations. Assigning the operator sets $\{\hat{A}_i\}_{i=1}^{L}$ and $\{\hat{a}_i\}_{i=1}^{2L+1}$ to the coarse and fine grids, respectively, using Eq.~\eqref{eq:creation}, we see that they inherit a similar relation:
\begin{equation}\label{eq:multigrid}
    \hat{A}_i^{\dagger} = \frac{1}{2\sqrt{2}}(\hat{a}_{2i-1}^{\dagger} + 2 \hat{a}_{2i}^{\dagger} + \hat{a}_{2i+1}^{\dagger}).
\end{equation}

The relation between the physical operators can further be translated into a mapping between the corresponding computational operators, $\{C_i^\dagger\}_{i=1}^L$ and $\{c_i^\dagger\}_{i=1}^{2L+1}$. Specifically, since $\mathrm{im}(W) = \mathbb{H}^{(L)} \subset \mathbb{H}^{(2L+1)}$, we can define a refinement map $R: \mathbb{H}_{c}^{(L)} \to \mathbb{H}_{c}^{(2L+1)}$ that lifts a coarse grid computational state to the physical space $\mathbb{H}$ and then maps it back to the fine grid computational space using the left inverse $W^+ = (W^\dagger W)^{-1} W^\dagger$, where $W = W^{(2L+1)}$. By composing these maps,
\begin{equation}\label{eq:refinement_defn}
    R = \left(W^{(2L+1)}\right)^{+} W^{(L)},
\end{equation}
we find that the refinement map is completely specified by the following relations:
\begin{subequations}
    \begin{align}
        R C_i^{\dagger} = \frac{1}{2\sqrt{2}}(c_{2i-1}^{\dagger} + 2c_{2i}^{\dagger} + c_{2i+1}^{\dagger}) R \label{eq:multigrid_comp} \\
        R \ket{\Omega}_c^{(L)} = \ket{\Omega}_c^{(2L+1)}\,.
    \end{align}
\end{subequations}
Note that $R$ is isometric with respect to the proper metric on the computational spaces, i.e.\ $R^\dagger \mathcal{N}^{(2L+1)} R = \mathcal{N}^{(L)}$, but not with respect to the trivial Euclidean inner product ($R^\dagger R \neq \openone$).

We can however decompose the refinement operator as a composition of two maps, $R = \mathcal{R} Q$. Here, the ``trivially isometric'' embedding $Q: \mathbb{H}_c^{(L)} \to \mathbb{H}_c^{(2L+1)}$ is defined by
\begin{align}
    Q C_i^\dagger & = c^\dagger_{2i} Q, & Q \ket{\Omega}_c^{(L)} = \ket{\Omega}_c^{(2L+1)},
\end{align}
whereas $\mathcal{R}: \mathbb{H}_c^{(2L+1)} \to \mathbb{H}_c^{(2L+1)}$ once again establishes a (invertible but non-unitary) single-particle transformation on the bigger computational Hilbert space, namely via
\begin{subequations}\label{eq:R_basis_defn}
    \begin{align}
        \mathcal{R} c^\dagger_{2i}          & = \frac{1}{2\sqrt{2}}({c}_{2i-1}^\dagger + 2{c}_{2i}^\dagger + {c}_{2i+1}^\dagger) \mathcal{R}, \\
        \mathcal{R} c^\dagger_{2i-1}        & = {c}_{2i-1}^\dagger \mathcal{R},                                                               \\
        \mathcal{R} \ket{\Omega}_c^{(2L+1)} & = \ket{\Omega}_c^{(2L+1)}.
    \end{align}
\end{subequations}

\subsubsection{Constructing the refinement map}
Let us now consider a many-body state in the MPS representation expanded in the coarse grid basis:
\begin{align}
    \ket{\Psi^{(L)}}_c & = \sum_{\{n_i\}} M^L_{\{n_i\}} \left(\prod_{i=1}^L \frac{1}{\sqrt{n_i!}} \, (C^{\dagger}_i)^{n_i} \right) \ket{\Omega}_c^{(L)},
\end{align}
\begin{equation*}
    \begin{tikzpicture}[baseline=(current bounding box.center), scale=0.7, semithick,
            tensor/.style={draw, rectangle, rounded corners=2pt, minimum width=1.8em, minimum height=1.4em, fill=white, font=\scriptsize}]

        \node[left] at (-1.2, 0) {$M^L = \;$};

        \node[mpstensor] (T1) at (0, 0) {$M^{(1)}$};
        \node[mpstensor] (T2) at (1.6, 0) {$M^{(2)}$};

        \node[mpstensor] (TL) at (5.0, 0) {$M^{(L)}$};

        \draw (T1.east) -- (T2.west);

        \draw (T2.east) -- (TL.west) node[midway, fill=white, inner sep=2pt, font=\small] {$\dots$};

        \draw (T1.south) -- ++(0,-0.5) node[below, font=\small] {$n_1$};
        \draw (T2.south) -- ++(0,-0.5) node[below, font=\small] {$n_2$};
        \draw (TL.south) -- ++(0,-0.5) node[below, font=\small] {$n_L$};
    \end{tikzpicture},
\end{equation*}
which has a corresponding physical state $\ket{\Psi} = W^{(L)} \ket{\Psi^{(L)}}_c$. We would now like to use the refinement procedure to express the same physical state in the refined grid basis, namely as $\ket{\Psi^{(2L+1)}}_c = R \ket{\Psi^{(L)}}_c$, where
\begin{align}
    \ket{\Psi^{(2L+1)}}_c & = \sum_{\{n_i\}} M^{2L+1}_{\{n_i\}} \left(\prod_{i=1}^{2L+1} \frac{1}{\sqrt{n_i!}} \, (c^{\dagger}_i)^{n_i} \right) \ket{\Omega}_c^{(2L+1)}.
\end{align}

Using the decomposition $R = \mathcal{R} Q$, we can first define the intermediate state $\ket{\tilde\Psi^{(2L+1)}}_c = Q \ket{\Psi^{(L)}}_c$ through the ``trivial'' embedding, which amounts to interleaving the tensors $M^L$ with tensors representing the Fock vacuum $\ket{\Omega}$ that act as identities on the virtual bonds. The resulting intermediate state $\ket{\tilde \Psi} \neq \ket{\Psi}$ is associated with the following computational state
\begin{align}
    \ket{\tilde\Psi^{(2L+1)}}_c & = \sum_{\{n_i\}} \tilde M^{2L+1}_{\{n_i\}} \left(\prod_{i=1}^{2L+1} \frac{1}{\sqrt{n_i!}} \, (\tilde c^{\dagger}_i)^{n_i} \right) \ket{\Omega}_c^{(2L+1)},
\end{align}
\begin{equation*}
    \begin{tikzpicture}[baseline=(current bounding box.center), scale=0.7, semithick,
            tensor/.style={draw, rectangle, rounded corners=2pt, minimum width=1.8em, minimum height=1.4em, fill=white, font=\scriptsize},
            vacuum/.style={draw=gray!80, text=black, rectangle, rounded corners=2pt, minimum width=1.8em, minimum height=1.4em, fill=gray!10, font=\scriptsize}]

        \node[left] at (-1.2, 0) {$\tilde M^{2L+1} = \;$};

        \node[vacuum] (V1) at (0, 0) {$\Omega$};
        \node[mpstensor] (B1) at (1.6, 0) {$M^{(1)}$};
        \node[vacuum] (V2) at (3.2, 0) {$\Omega$};
        \node[mpstensor] (BL) at (5.6, 0) {$M^{(L)}$};
        \node[vacuum] (VL_vac) at (7.2, 0) {$\Omega$};

        \draw (V1.east) -- (B1.west);
        \draw (B1.east) -- (V2.west);
        \draw (V2.east) -- (BL.west) node[midway, fill=white, inner sep=1pt, font=\small] {$\dots$};
        \draw (BL.east) -- (VL_vac.west);

        \draw (V1.south) -- ++(0,-0.5) node[below, font=\small] {$n_1$};
        \draw (B1.south) -- ++(0,-0.5) node[below, font=\small] {$n_2$};
        \draw (V2.south) -- ++(0,-0.5) node[below, font=\small] {$n_3$};
        \draw (BL.south) -- ++(0,-0.5) node[below, font=\small] {$n_{2L}$};
        \draw (VL_vac.south) -- ++(0,-0.5) node[below, font=\small] {$n_{2L+1}$};
    \end{tikzpicture},
\end{equation*}
\begin{equation*}
    \begin{tikzpicture}[baseline=(current bounding box.center), scale=0.7, semithick,
            vacuum/.style={draw=gray!80, text=black, rectangle, rounded corners=2pt, minimum width=1.8em, minimum height=1.4em, fill=gray!10, font=\scriptsize}]

        \node[vacuum] (V) at (0, 0) {$\Omega$};

        \draw (V.west) -- ++(-0.6,0) node[left, font=\small] {$i$};
        \draw (V.east) -- ++(0.6,0) node[right, font=\small] {$j$};
        \draw (V.south) -- ++(0,-0.5) node[below, font=\small] {$n$};

        \node[right] at (1.2, 0) {$\ \ = \ \delta_{ij} \delta_{n0}.$};
    \end{tikzpicture}
\end{equation*}

Next, the nontrivial action of $\mathcal{R}$ can once again be represented as an MPO. We directly employ the particle-flow picture introduced in Sec.~\ref{sec:W_construction} instead of performing the full algebraic manipulation of the multinomial expansion. We start by noting that the refinement process is driven entirely by the distribution of particles from the even sites due to the bipartite structure of the transformation  Eq.~\eqref{eq:R_basis_defn}. At an even site $j = 2i$, the $n_j$ particles are split into three paths: $k_{j, +1}$ particles move to the left, $k_{j, -1}$ move to the right, and $k_{j, 0}$ stay on the same site. We identify these outgoing paths directly with the adjacent virtual bonds, $Q_{j-1} = k_{j, +1}$ and $Q_j = k_{j, -1}$, where $Q_j$ is simply the scalar particle flow crossing the link between $j$ and $j+1$ as defined previously. From particle number conservation, it follows that
\begin{equation}
    n_j' = k_{j, 0} = n_j - Q_{j-1} - Q_j,
\end{equation}
On the other hand, the odd-indexed sites $j = 2i-1$ act as sinks that simply accumulate the particles from their left and right neighbors, and combines them with those already present:
\begin{equation}
    n_j' = n_j + Q_{j-1} + Q_j.
\end{equation}
When acting on a state created by the trivial embedding $Q$, such as $\ket{\tilde\Psi^{(2L+1)}}_c$, only $n_j=0$ will contribute.

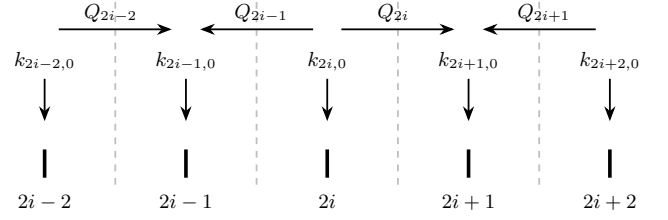
\begin{figure}[t]
    \centering
    \resizebox{\columnwidth}{!}{
        \begin{tikzpicture}[
                xscale=2.2,
                yscale=1.1,
                >=Stealth,
                site line/.style={ultra thick},
                boundary line/.style={gray!50, dashed, thick},
                flow box/.style={draw=none, fill=gray!20, fill opacity=0.8, text opacity=1, rounded corners=3pt, inner sep=3pt, font=\scriptsize},
                label_left/.style={font=\small, anchor=south, xshift=-23pt, inner sep=3pt},
                label_right/.style={font=\small, anchor=south, xshift=23pt, inner sep=3pt},
                label_down/.style={font=\small, anchor=south, xshift=0pt, inner sep=3pt}
            ]
            \def\yoffset{0.4}

            \foreach \x in {-0.5, 0.5, 1.5, 2.5} {
                    \draw[boundary line] (\x, -0.1) -- (\x, 2.5);
                }

            \foreach \x/\label in {-1/2i-2, 0/2i-1, 1/2i, 2/2i+1, 3/2i+2} {
                    \draw[site line] (\x, 0) -- (\x, 0.4);
                    \node[below=4pt, font=\small] at (\x, 0) {$\label$};
                }

            \begin{scope}[yshift=\yoffset cm]
                \foreach \x/\label in {-1/2i-2, 0./2i-1, 1./2i, 2./2i+1, 3./2i+2} {
                        \draw[->, thick] (\x, 1.) node[label_down] {$k_{\label, 0}$} -- (\x, 0.4);
                    }

                \draw[->, thick] (2.9, 1.7) node[label_left] {$Q_{2i+1}$} -- (2.1, 1.7);
                \draw[->, thick] (1.1, 1.7) node[label_right] {$Q_{2i}$} -- (1.9, 1.7);
                \draw[->, thick] (0.9, 1.7) node[label_left] {$Q_{2i-1}$} -- (0.1, 1.7);
                \draw[->, thick] (-0.9, 1.7) node[label_right] {$Q_{2i-2}$} -- (-0.1, 1.7);

            \end{scope}
        \end{tikzpicture}
    }
    \caption{Visual schematic of the particle-flow picture for the refinement map. The indices $k_{i,0}$ count the particles remaining at site $i$, while the virtual indices $Q_i$ count the particles moving to neighboring sites.}
    \label{fig:particle_flow_refinement}
\end{figure}

These rules determine the local MPO tensors, where the $Q_j$ serve as the virtual bond and the entries are determined by the multinomial coefficients as before:
\begin{equation}
    \begin{aligned}
        \begin{tikzpicture}[baseline=(current bounding box.center), scale=0.45, semithick]
            \node[mpotensor] (T) at (0,0) {\small $\mathcal{R}^{(j)}$};
            \draw (T.north) -- ++(0,0.7) node[above] {\scriptsize $n_j$};
            \draw (T.west) -- ++(-0.7,0) node[left] {\scriptsize $Q_{j-1}$};
            \draw (T.south) -- ++(0,-0.7) node[below] {\scriptsize $n_j'$};
            \draw (T.east) -- ++(0.7,0) node[right] {\scriptsize $Q_j$};

        \end{tikzpicture}
         & = \delta_{n_j', \, n_j - Q_{j-1} - Q_j} \sqrt{\frac{n_j'!}{n_j!}} \\[-5.5ex]
         & \times \binom{n_j}{Q_{j-1}, n_j', Q_j} \; 2^{n_j' - 3n_j/2}
    \end{aligned}
\end{equation}
for even $j$, and
\begin{equation}
    \begin{aligned}
        \begin{tikzpicture}[baseline=(current bounding box.center), scale=0.45, semithick]
            \node[altmpotensor] (T) at (0,0) {\small $\mathcal{R}^{(j)}$};
            \draw (T.north) -- ++(0,0.7) node[above] {\scriptsize $n_j$};
            \draw (T.west) -- ++(-0.7,0) node[left] {\scriptsize $Q_{j-1}$};
            \draw (T.south) -- ++(0,-0.7) node[below] {\scriptsize $n_j'$};
            \draw (T.east) -- ++(0.7,0) node[right] {\scriptsize $Q_j$};

        \end{tikzpicture}
         & =\delta_{n_j', \, n_j + Q_{j-1} + Q_j} \, \sqrt{\frac{n_j'!}{n_j!}}
    \end{aligned}
\end{equation}
for odd $j$. As no particles interchange positions in this case, no additional sign factors are needed when this construction is applied to fermionic systems.

Note that, alternatively, the refinement map $R$ can be directly interpreted as an atypical MPO as shown below for $L=5$ (with $n_{2i+1} = 0$ and $n_{2i} = \tilde n_i$).
\begin{equation*}
    \begin{tikzpicture}[baseline=(current bounding box.center), scale=0.8, semithick,
            tensor/.style={draw, rectangle, rounded corners=2pt, minimum width=2.2em, minimum height=1.6em, fill=white, font=\scriptsize}]

        \node[altmpotensor] (T1) at (0, 0) {$R^{(1)}$};
        \node[mpotensor] (T2) at (2, 0) {$R^{(2)}$};
        \node[altmpotensor] (T3) at (4, 0) {$R^{(3)}$};
        \node[mpotensor] (T4) at (6, 0) {$R^{(4)}$};
        \node[altmpotensor] (T5) at (8, 0) {$R^{(5)}$};

        \draw (T1.east) -- (T2.west);
        \draw (T2.east) -- (T3.west);
        \draw (T3.east) -- (T4.west);
        \draw (T4.east) -- (T5.west);

        \foreach \i in {1,2,3,4,5} {
                \draw (T\i.south) -- ++(0,-0.6) node[below, font=\small] {$n_{\i}'$};
            }

        \draw (T2.north) -- ++(0,0.6) node[above, font=\small] {$\tilde n_1$};
        \draw (T4.north) -- ++(0,0.6) node[above, font=\small] {$\tilde n_2$};
    \end{tikzpicture} \, ,
\end{equation*}
The construction presented in this section is however more practical for implementation using standard numerical MPS/MPO libraries. In either case, this results in an MPO representation with bond dimension $\chi = n_c + 1$ and a bound on the resulting local occupations, $0 \leq n_j' \leq 2n_c$. In practice, we enforce a fixed cutoff $n_c$ on the refined state, but this does not introduce significant error since the population of the modes that we truncate away naturally vanishes in the continuum limit. Additionally, applying the MPO increases the MPS bond dimension by a factor of $\chi$. We can choose to compress the state back to its initial bond dimension after refinement and observe that this typically introduces $< 5\%$ energy error for physically relevant states. Despite these practical considerations, we find that the refined state provides a very good initial state that accelerates subsequent optimization on the refined grid.

\section{Results}\label{sec:results}
In this section, we demonstrate the formalism using a single-species system of bosons with contact interactions as described by the Hamiltonian in Eq.~\eqref{eq:LiebLiniger}. To provide a point of reference, we compare the performance of the finite-element MPS (FE-MPS) against a standard finite-difference MPS (FD-MPS) treatment (see \appref{app:finite_diff}). Unless specified otherwise, all further computations utilize a particle number cutoff $n_c=2$ for FE-MPS and FD-MPS. The optimization on a specific grid is performed starting from an initial state, then running a single sweep of 2-site DMRG to increase the bond dimension, followed by 1-site DMRG sweeps until the gradient norm of the energy drops below an absolute tolerance of $10^{-5}$. Note that we work in the grand-canonical ensemble and tune the chemical potential $\mu$ to reach a target particle number $N$. However, this formalism naturally accommodates the utilization of  $U(1)$-symmetric tensors to enforce particle number conservation as well.

\subsection{Benchmarks}

\begin{figure}[t]
    \centering
    \includegraphics[width=\columnwidth]{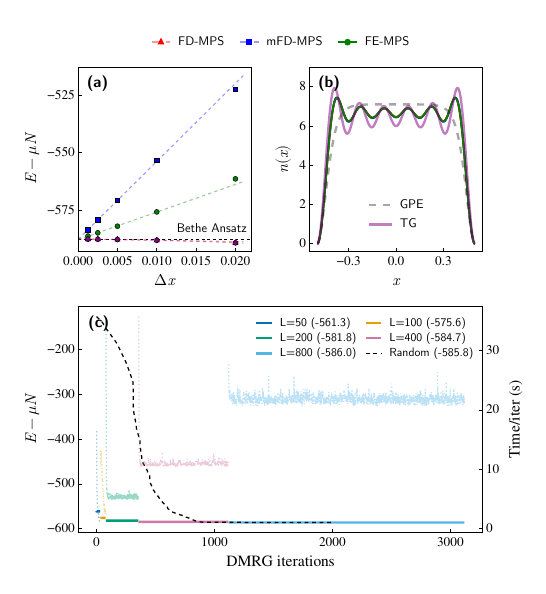}
    \caption{Ground state properties and convergence behavior for $N=6$ particles in an infinite well of size $l=1$ with $g=50$ at bond dimension $D=10$. (a) Scaling of the ground state energy error with grid spacing $\Delta x$ for FD-MPS (red, $n_c=2$), modified FD-MPS (blue, $n_c=1$, see \appref{app:finite_diff}), and FE-MPS (green, $n_c=2$). The absolute error of the continuum limit relative to the Bethe ansatz energy is due to small $D$. (b) Comparison of the spatial density $n(x)$ at the finest resolution ($L=800$) against Gross-Pitaevskii (GPE) and Tonks-Girardeau (TG) limits. (c) Energy convergence during DMRG optimization of FE-MPS, demonstrating the performance of the multigrid optimization scheme compared to a standard random initialization. The dotted lines indicate the computational time spent per DMRG iteration.}
    \label{fig:infwell_conv}
\end{figure}

\begin{figure}[t]
    \centering
    \includegraphics[width=\columnwidth]{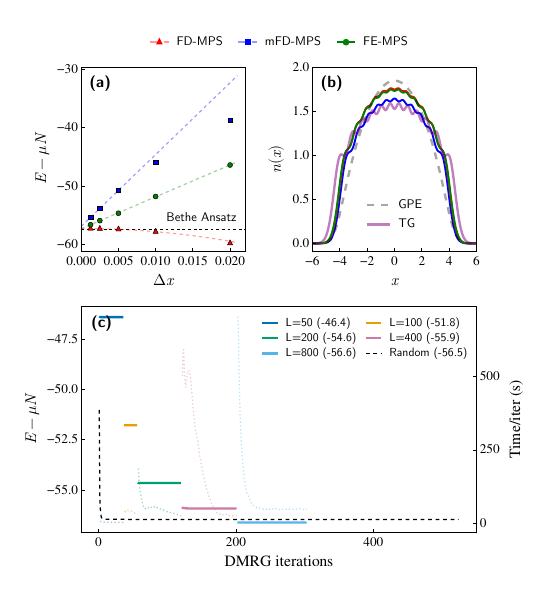}
    \caption{Ground state properties and convergence behavior as in Fig.~\ref{fig:infwell_conv}, but for $N=12$ particles in a harmonic trap with $\omega = 1$ and $g=10$ at bond dimension $D=15$. (a) Energy convergence with $\Delta x$ where we compare against the Bethe ansatz solution using a local density approximation (see ~\appref{app:bethe_ansatz}). (b) Comparison of the spatial density $n(x)$ at $L=800$. (c) Energy convergence during DMRG optimization of FE-MPS.}
    \label{fig:qho_conv}
\end{figure}

To evaluate the convergence properties of the FE-MPS, we consider two benchmark cases: $N=6$ particles in an infinite square well (Fig.~\ref{fig:infwell_conv}) and $N=12$ particles in a harmonic trap (Fig.~\ref{fig:qho_conv}).

The energy convergence in panels (a) reflects a fundamental trade-off: while standard FD-MPS achieves $O(\Delta x^2)$ convergence with the smallest absolute errors, it lacks variationality and may even exhibit non-monotonic behavior, as we will demonstrate in the subsequent section. Conversely, FE-MPS provides a strictly variational upper bound that approaches the continuum limit monotonically, but at an observed rate of $O(\Delta x)$ for this case. As we illustrate in Appendix~\ref{app:convergence}, this linear scaling is due to the presence of cusps in the many-body wavefunction induced by the $\delta$-interaction, which limits the expressivity of the piecewise-linear basis. The modified FD-MPS variant, designed for compatibility with a cMPS-based multigrid scheme \cite{ganahl_continuous_2018}, also appears to converge variationally with a rate $O(\Delta x)$. However, unlike FE-MPS, the variational nature is in this case merely an empirical observation rather than a formal statement, and it comes at the cost of much larger absolute errors for comparable grid sizes.

\begin{figure}[t]
    \centering
    \includegraphics[width=\columnwidth]{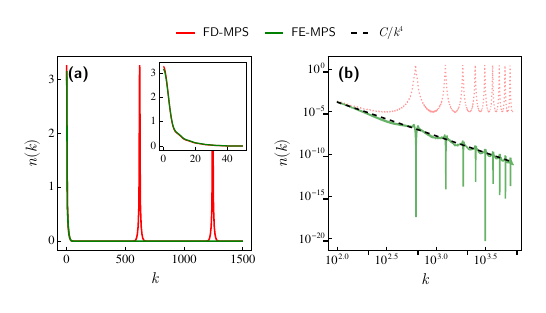}
    \caption{Momentum distribution $n(k)$ (see \appref{app:momentum_dist}) for the parameters in Fig.~\ref{fig:infwell_conv} at $L=100$. (a) Linear scale comparison highlighting the agreement between FD-MPS (red) and FE-MPS (green) on IR behavior. The extended $k$-range demonstrates periodic aliasing artifacts inherent to the FD representation, which are suppressed with the FE basis. Inset: Zoom-in on the low-$k$ regime, showing the identical resolution of the bulk density peak. (b) Log-log scale illustrating the UV behavior. FE-MPS recovers the physical $k^{-4}$ power-law tail \cite{Rigol2015}, while FD-MPS plateaus with periodic behavior. The black dashed line represents the theoretical asymptotic limit $C/k^4$, where the contact $C = 4g E_{\text{int}}$ is calculated independently using the interaction energy \cite{Rigol2015}.}
    \label{fig:tan_contact}
\end{figure}

Panels (b) of Figs.~\ref{fig:infwell_conv} and \ref{fig:qho_conv} demonstrate that the spatial density of FE-MPS converges more rapidly than the ground-state energy and is in good agreement with FD-MPS. A distinct advantage of FE-MPS is that all observables are represented directly in the continuum by construction, whereas FD-MPS yields discrete points that require interpolation for further use. For instance, FE-MPS naturally captures ultraviolet (UV) properties such as Tan's contact \cite{TAN20082952} as shown in Fig.~\ref{fig:tan_contact}.

Finally, the DMRG trajectories in panels (c) illustrate the utility of a multigrid optimization scheme. The sharp spikes in computational time per DMRG iteration mark the grid refinement steps, where the algorithm briefly requires more time to recompute the new environment tensors before settling into a long plateau while resolving finer spatial features.

While the total runtime of the multigrid approach is often comparable to a direct optimization on the finest grid starting from a random initialization, it offers the distinct advantage of providing converged intermediate results essentially for free. Hence, the data points required for a systematic continuum extrapolation are obtained within a single optimization run without requiring several independent simulations. Furthermore, progressive refinement serves as a more robust optimization strategy as it consistently locates a lower ground state energy, thereby avoiding the local minima that often trap random initial states \cite{dolfi_multigrid_2012} and ensuring monotonic convergence.

\subsection{Gaussian barrier}
To demonstrate a scenario where FE-MPS offers a distinct practical advantage over FD-MPS, we consider a system featuring multiple length scales that are difficult to resolve without high grid resolutions. Specifically, we introduce a narrow Gaussian barrier at the center of a harmonic trap:
\begin{equation}
    V(x) = \frac{1}{2}\omega^2 x^2 + \Omega e^{-x^2/2\sigma^2}.
\end{equation}
We set the barrier width to $\sigma = 0.01 \; l_h$ (with $l_h = \sqrt{\hbar/m\omega} = 1$ here), which is significantly smaller than the typical grid spacings $\Delta x$ that we will consider. This potential serves as a model for a Dirac delta impurity that may be realized experimentally using a tightly focused laser beam or an impurity atom whose interaction strength is tuned by a Feshbach resonance.

\begin{figure*}[t]
    \hspace*{0.75cm}
    \centering
    \includegraphics[width=0.98\textwidth]{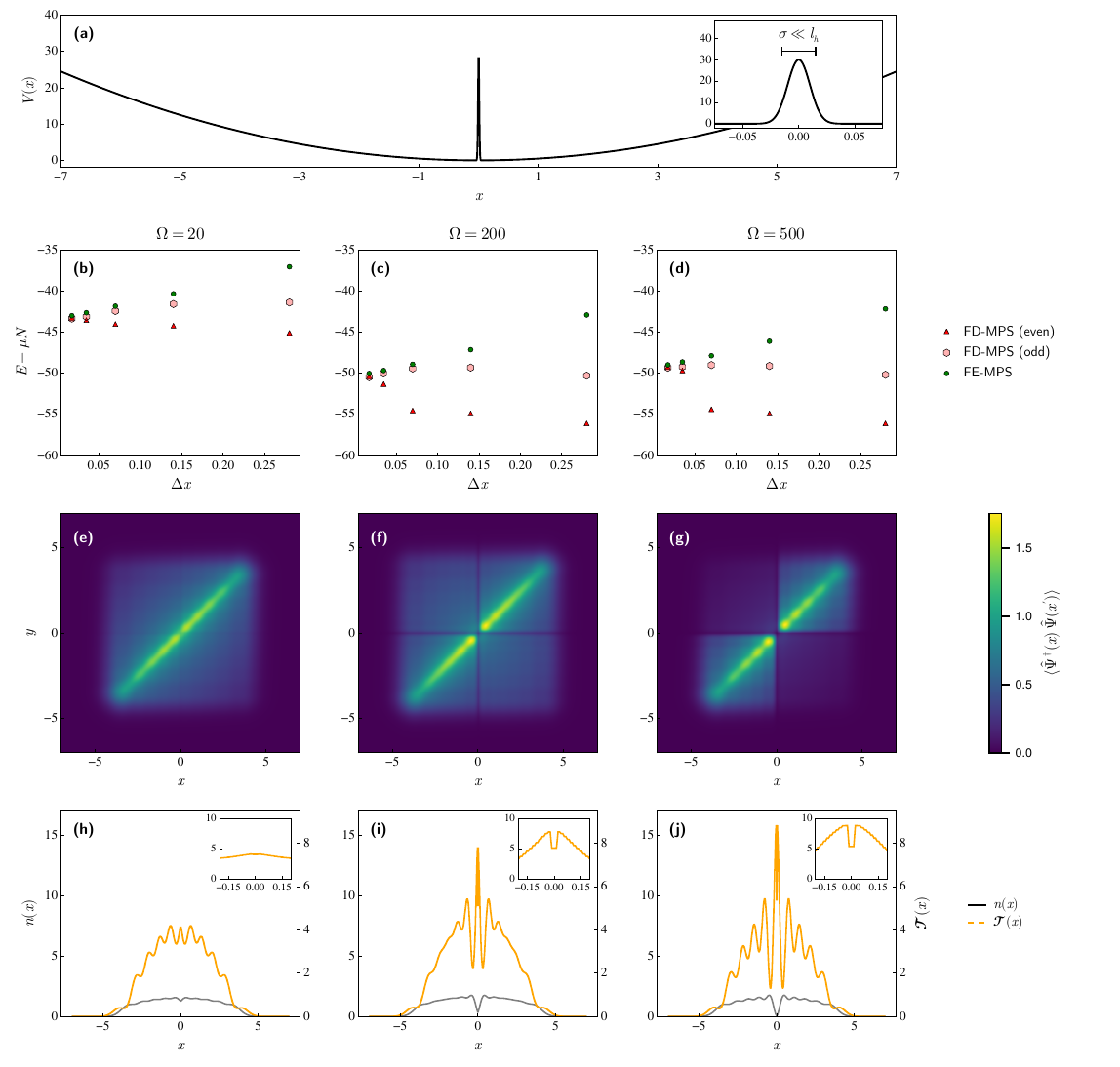}
    \caption{Ground state properties and convergence behavior of $N=10$ particles in a harmonic trap with a Gaussian impurity of width $\sigma=0.01 \; l_h$ with $g=10$ at bond dimension $D=20$. (a) Potential energy profile $V(x)$ showing the Gaussian impurity. (b)--(d) Energy convergence as a function of grid spacing $\Delta x$ for barrier heights $\Omega \in \{20, 200, 500\}$. FD-MPS $(n_c = 5)$ results are shown for even grid sizes $L\in[100, 200, 400, 800]$ (red) and odd grid sizes $L\in[101, 201, 401, 801]$ (pink), highlighting the non-monotonic convergence. (e)--(g) Real-space single-particle density matrices $\langle \hat{\Psi}^{\dagger}(x) \hat{\Psi}(x^\prime) \rangle$ resolved for the $L=800$ FE-MPS. (h)--(j) Corresponding particle densities, $n(x) = \langle\hat \Psi^{\dagger}(x)\hat\Psi(x)\rangle$ and kinetic energy densities, $\mathcal{T}(x)=1/2 \cdot\langle\partial_x\hat\Psi^{\dagger}(x)\partial_x\hat\Psi(x)\rangle$, resolved for the $L=800$ FE-MPS.}
    \label{fig:qho_impurity}
\end{figure*}

As shown in panels (b)–(d) of Fig.~\ref{fig:qho_impurity}, FE-MPS maintains a stable, monotonic $\mathcal{O}(\Delta x)$ convergence that respects the variational bound across the full range of barrier strengths $\Omega$. In contrast, the standard FD-MPS exhibits severely degraded convergence and extreme sensitivity to the choice of grid points. Since FD schemes rely on point-wise sampling, the result depends on whether a grid node happens to align with the narrow barrier peak, leading to non-monotonic convergence that make it cumbersome to perform a systematic extrapolation to the continuum limit. The finite-element basis naturally avoids this issue by integrating the potential over the spatial support of the tent functions, ensuring that sub-grid features are variationally accounted for even at coarse resolutions.

The physical impact of the barrier is captured in the single-particle density matrix (SPDM) in panels (e)–(g). As $\Omega$ increases, we observe the emergence of a block-diagonal structure, signifying the destruction of spatial coherence and the fragmentation of the gas into two isolated wells.

This fragmentation is further highlighted by the local kinetic energy density shown in panels (h)–(j), which develops sharp spikes at the center due to the large spatial gradients forced by the density depletion at the location of the barrier. Ultimately, this demonstrates that FE-MPS robustly captures both the local densities and the long-range correlations of continuum systems, providing a reliable numerical framework in settings where standard finite-difference schemes fall short.

\section{Outlook}\label{sec:outlook}

While we have primarily demonstrated this framework on UV-finite continuum theories using a FE discretization, the underlying methodology for handling basis non-orthogonality within the MPS formalism is much more general. A natural future direction is applying the generalized DMRG scheme to quantum chemistry or solid-state physics settings, where non-orthogonal basis sets (e.g., Gaussian-type or numerical atomic orbitals) are regularly employed. Extending this framework in the context of the time-dependent variational principle (TDVP) would similarly enable the simulation of non-equilibrium dynamics in these systems.

Focusing specifically on FE methods, several natural avenues remain to be explored as well. For instance, one could implement adaptive $h$-refinement using non-uniform grids to resolve sharp density gradients without requiring a global increase in resolution. Additionally, we may use higher-order elements with larger local support (i.e, higher bandwidth $R$ of the overlap matrix $N$) to recover faster energy convergence, albeit at the cost of a larger many-body overlap tensor. Alternatively, one could introduce variational degrees of freedom in the single-particle orbitals themselves. Finally, the utility of FE methods are most evident in higher-dimensional settings with non-trivial geometries, where one may conceivably utilize tree tensor networks or projected entangled-pair states to capture the resulting physical connectivity. We leave these developments for future work.

\medskip
{\noindent \textit{Note.} After the initial posting of this manuscript, we became aware of complementary work by Kaplan and Tilloy \cite{kaplan2026}, which uses a local and orthogonal basis of Daubechies scaling functions to discretize field theories. This approach also naturally admits refinement which, in combination with imposing translation invariance, seems to be effective in further lowering the energy. However, the required smoothness of the basis necessitates a wide spatial support (i.e, $D6$ wavelets), yielding larger Hamiltonian MPO bond dimensions ($\chi = 702$) than the tent function basis ($\chi = 81$ for $n_c=2$). Both our approaches thus offer distinct and complementary benefits.}

\section*{Code and Data Availability}
The numerical implementation of the scheme is written in Julia v1.11 \cite{julia_lang} and utilizes the \texttt{TensorKit.jl} \cite{tensorkit}, \texttt{MPSKit.jl} \cite{mpskit}, and \texttt{KrylovKit.jl} \cite{krylovkit} frameworks. Data management and visualization are handled by \texttt{DrWatson.jl} \cite{Datseris2020} and \texttt{Makie.jl} \cite{makie}, respectively. The source code and analysis scripts are maintained on GitHub \cite{TentMPS_pkg, TentMPS_scripts}, and a self-contained archive including the raw data is available on Zenodo \cite{TentMPS_data}.

\begin{acknowledgements}
    We acknowledge useful discussions with Atsushi Ueda, Frank Verstraete and Lukas Devos. This work is supported by the FWO research grant G0C5123N.
\end{acknowledgements}

\appendix
\section{Constructing the Hamiltonian MPO}
\label{app:hamiltonian_specifics}
In this Appendix, we construct the Hamiltonian in Eq.~\eqref{eq:LiebLiniger} as an MPO to facilitate the variational optimization of the ground state using the generalized DMRG algorithm developed in Sec.~\ref{sec:generalized_dmrg}. As mentioned in the main text, since the tent functions $\phi_i(x)$ only have local support, all matrix elements vanish unless the site indices are identical or adjacent. Below we compute these non-vanishing elements arising from each term of the continuum Hamiltonian in Eq.~\eqref{eq:LiebLiniger}.

\subsection*{Kinetic energy}
The derivative of the basis functions Eq.~\eqref{eq:tent_fn} is a piecewise constant step function:

\begin{equation}
    \partial_x\phi_i(x) =
    \begin{cases}
        1/h, & x \in [x_{i-1}, x_i], \\
        -1/h & x \in (x_i, x_{i+1}], \\
        0,   & \text{otherwise,}
    \end{cases}
\end{equation}
with $h = \sqrt{2 \Delta x^3 / 3}$.
Starting from the kinetic energy operator in its manifestly Hermitian and positive definite formulation,
\begin{equation}
    \hat{T} = \frac{1}{2} \int_{-\ell}^{\ell} dx \, \partial_x\Psi^{\dagger}\partial_x\Psi,
\end{equation}
we obtain the matrix elements:

\begin{subequations}
    \begin{align}
        t^k_{i,i}   & = \frac{1}{2}\int_{-\ell}^{+\ell} \dd x \, (\partial_x\phi_{i}(x))^2 = \frac{3}{2 \Delta x^2}                    \\
        t^k_{i,i+1} & = \frac{1}{2}\int_{-\ell}^{+\ell} \dd x \, \partial_x\phi_i(x) \partial_x\phi_{i+1}(x) = -\frac{3}{4 \Delta x^2}
    \end{align}
\end{subequations}

\subsection*{Chemical potential}
The chemical potential matrix elements are equivalent to that of the single-particle overlap matrix $N$:
\begin{subequations}
    \begin{align}
        t^c_{i,i}   & = \int_{-\ell}^{+\ell} \dd x \, \phi_{i}(x)^2 = 1                     \\
        t^c_{i,i+1} & = \int_{-\ell}^{+\ell} \dd x \, \phi_i(x) \phi_{i+1}(x) = \frac{1}{4}
    \end{align}
\end{subequations}

\subsection*{Harmonic potential}
For the specific case $V(x) = \frac{1}{2}x^2$, the matrix elements depend on the site coordinates $x_i$ as:
\begin{subequations}
    \begin{align}
        t^v_{i,i}   & = \frac{1}{2}\int_{-\ell}^{+\ell} \dd x \, x^2 \phi_i^2(x) \nonumber               \\
                    & = \frac{1}{20} (10 x_i^2 + \Delta x^2)                                             \\
        t^v_{i,i+1} & = \frac{1}{2} \int_{-\ell}^{+\ell} \dd x \, x^2 \phi_{i}(x)\phi_{i+1}(x) \nonumber \\
                    & = \frac{1}{80}(10 x_i^2 + 10 \Delta x x_i + 3\Delta x^2).
    \end{align}
\end{subequations}
For a generic potential, these may be computed to machine precision using numerical Gaussian quadrature.

\subsection*{Two-body interactions}
For the contact interaction we have the following contributions:
\begin{subequations}
    \begin{align}
        U_{iiii} & =  g\int_{-\ell}^{+\ell} dx \, \phi_{i}(x)^4 = g\frac{9}{10 \Delta x}            \\
        U_{iiij} & = g\int_{-\ell}^{+\ell} dx \, \phi_i^3(x) \phi_{j}(x) = g\frac{9}{80 \Delta x}   \\
        U_{iijj} & = g\int_{-\ell}^{+\ell} dx \, \phi_i^2(x) \phi_{j}^2(x) = g\frac{3}{40 \Delta x}
    \end{align}
\end{subequations}
where $j=i+1$ and the only non-vanishing elements are for the permutations of these indices.

\subsection*{Full Hamiltonian and its MPO representation}
Putting all of this together, we arrive at the Hamiltonian:
\begin{align}
    \hat{H} ={} & \sum_{i} \bigl[ t_{i,i} (\hat{b}^i)^{\dagger} \hat{b}^i + (t_{i,i+1} (\hat{b}^i)^{\dagger} \hat{b}^{i+1} + t_{i+1,i} (\hat{b}^{i+1})^{\dagger} \hat{b}^i) \bigr] \nonumber \\
    + \;        & \sum_{i} \Bigl[ U_{iiii} \fourb{i,i,i,i} \nonumber                                                                                                                         \\
                & \quad + (U_{iiij} + U_{iiji}) \fourb{i,i,i,j} \nonumber                                                                                                                    \\
                & \quad + (U_{ijii} + U_{jiii}) \fourb{j,i,i,i} \nonumber                                                                                                                    \\
                & \quad + (U_{jjji} + U_{jjij}) \fourb{j,j,j,i} \nonumber                                                                                                                    \\
                & \quad + (U_{jijj} + U_{ijjj}) \fourb{i,j,j,j} \nonumber                                                                                                                    \\
                & \quad + (U_{ijij} + U_{jiij} + U_{jiji} + U_{ijji})\fourb{i,j,j,i} \nonumber                                                                                               \\
                & \quad + U_{iijj} \fourb{i,i,j,j} + U_{jjii} \fourb{j,j,i,i} \Bigr]_{j=i+1},
\end{align}
where the contact interaction generates 8 distinct terms.

This is represented as an MPO of bond-dimension 9 by specifying the Jordan block structure \cite{schollwockDensitymatrixRenormalizationGroup2011} at each site,
\begin{align}
    \hat{H}^{(i)} =
    \begin{cases}
        \left[ \begin{array}{ccc} \mathbb{I} & \mathbf{L}^{(i)} & C^{(i)} \end{array} \right] & i = 1, \\[2ex]
        \left[ \begin{array}{cc|c}
                       \mathbb{I} & \mathbf{L}^{(i)} & C^{(i)}          \\ \hline
                       0          & 0                & \mathbf{R}^{(i)} \\
                       0          & 0                & \mathbb{I}
                   \end{array} \right]                            & 1 < i < L,                             \\[6ex]
        \left[ \begin{array}{c} C^{(i)} \\ \mathbf{R}^{(i)} \\ \mathbb{I} \end{array} \right] & i = L,
    \end{cases}
\end{align}
with the sub-blocks given by
\begin{subequations}
    \begin{align}
        C^{(i)}          & = U_{iiii} (\hat{b}^{\dagger})^2 (\hat{b})^2 + t_{ii} \hat{b}^{\dagger} \hat{b} \\
        \mathbf{L}^{(i)} & = \left[
            \begin{array}{r l}
                (U_{iiij} + U_{iiji})                                                                               & (\hat{b}^\dagger)^2 \hat{b} + t_{ij} \hat{b}^{\dagger} \\[0.5ex]
                (U_{ijii} + U_{jiii})                                                                               & \hat{b}^\dagger \hat{b}^2 + t_{ji} \hat{b}             \\[0.5ex]
                (U_{jjji} + U_{jjij})                                                                               & \hat{b}                                                \\[0.5ex]
                (U_{jijj} + U_{ijjj})                                                                               & \hat{b}^\dagger                                        \\[1ex]
                \left( \begin{aligned} U_{ijij} + U_{jiij} \\[0.5ex] + \, U_{jiji} + U_{ijji} \end{aligned} \right) & \hat{b}^{\dagger} \hat{b}                              \\[1.5ex]
                U_{iijj}                                                                                            & (\hat{b}^\dagger)^2                                    \\
                U_{jjii}                                                                                            & (\hat{b})^2
            \end{array}
        \right]^T                                                                                          \\
        \mathbf{R}^{(i)} & = \left[
            \begin{array}{c}
                \hat{b}                     \\
                \hat{b}^{\dagger}           \\
                (\hat{b}^\dagger)^2 \hat{b} \\
                \hat{b}^\dagger \hat{b}^2   \\
                \hat{b}^{\dagger}\hat{b}    \\
                (\hat{b})^2                 \\
                (\hat{b}^{\dagger})^2
            \end{array}
            \right],
    \end{align}
\end{subequations}
where again $j=i+1$ and we drop the indices on the operators.

In order to construct the effective Hamiltonian in the computational basis as shown in Eq.~\eqref{eq:LiebLiniger_comp}, we modify the MPO as described in Eq.~\eqref{eq:Z_normal_ordered}. Given the MPO representation of the many-body overlap in Eq.~\eqref{eq:overlap_mpo}, the Jordan block form follows as
\begin{align}
    \mathcal{H}^{(i)} =
    \begin{cases}
        \left[ \begin{array}{ccc} \mathcal{N}^{(i)} & \mathbf{L}^{(i)} & C^{(i)} \end{array} \right] & i = 1, \\[2ex]
        \left[ \begin{array}{cc|c}
                       \mathcal{N}^{(i)} & \mathbf{L}^{(i)} & C^{(i)}           \\ \hline
                       0                 & 0                & \mathbf{R}^{(i)}  \\
                       0                 & 0                & \mathcal{N}^{(i)}
                   \end{array} \right]                           & 1 < i < L,                              \\[6ex]
        \left[ \begin{array}{c} C^{(i)} \\ \mathbf{R}^{(i)} \\ \mathcal{N}^{(i)} \end{array} \right] & i = L,
    \end{cases}
\end{align}
now with
\begin{subequations}
    \begin{align}
        C^{(i)}          & = U_{iiii} (c^{\dagger})^2 \mathcal{N}^{(i)} c^2 + t_{ii} c^{\dagger} \mathcal{N}^{(i)} c \\
        \mathbf{L}^{(i)} & = \left[
            \begin{array}{r l}
                (U_{iiij} + U_{iiji})                                                                                 & (c^\dagger)^2 \mathcal{N}^{(i)} c + t_{ij} c^{\dagger} \mathcal{N}^{(i)} \\[1ex]
                (U_{ijii} + U_{jiii})                                                                                 & c^\dagger \mathcal{N}^{(i)} c^2 + t_{ji} \mathcal{N}^{(i)} c             \\[1ex]
                (U_{jjji} + U_{jjij})                                                                                 & \mathcal{N}^{(i)} c                                                      \\[1ex]
                (U_{jijj} + U_{ijjj})                                                                                 & c^\dagger \mathcal{N}^{(i)}                                              \\[1ex]
                \left( \begin{aligned} &U_{ijij} + U_{jiij} \\[0.5ex] &+ \, U_{jiji} + U_{ijji} \end{aligned} \right) & c^{\dagger} \mathcal{N}^{(i)} c                                          \\[2.5ex]
                U_{iijj}                                                                                              & (c^\dagger)^2 \mathcal{N}^{(i)}                                          \\
                U_{jjii}                                                                                              & \mathcal{N}^{(i)} c^2
            \end{array}
        \right]^T                                                                                                    \\
        \mathbf{R}^{(i)} & = \left[
            \begin{array}{c}
                \mathcal{N}^{(i)} c               \\
                c^{\dagger} \mathcal{N}^{(i)}     \\
                (c^\dagger)^2 \mathcal{N}^{(i)} c \\
                c^\dagger \mathcal{N}^{(i)} c^2   \\
                c^{\dagger} \mathcal{N}^{(i)} c   \\
                \mathcal{N}^{(i)} c^2             \\
                (c^{\dagger})^2 \mathcal{N}^{(i)}
            \end{array}
            \right],
    \end{align}
\end{subequations}
resulting in an effective bond dimension of $9(n_c +1)^2$ for a particle cutoff $n_c$.

\section{Convergence properties}
\label{app:convergence}

\begin{figure*}[t]
    \centering
    \includegraphics[width=0.8\textwidth]{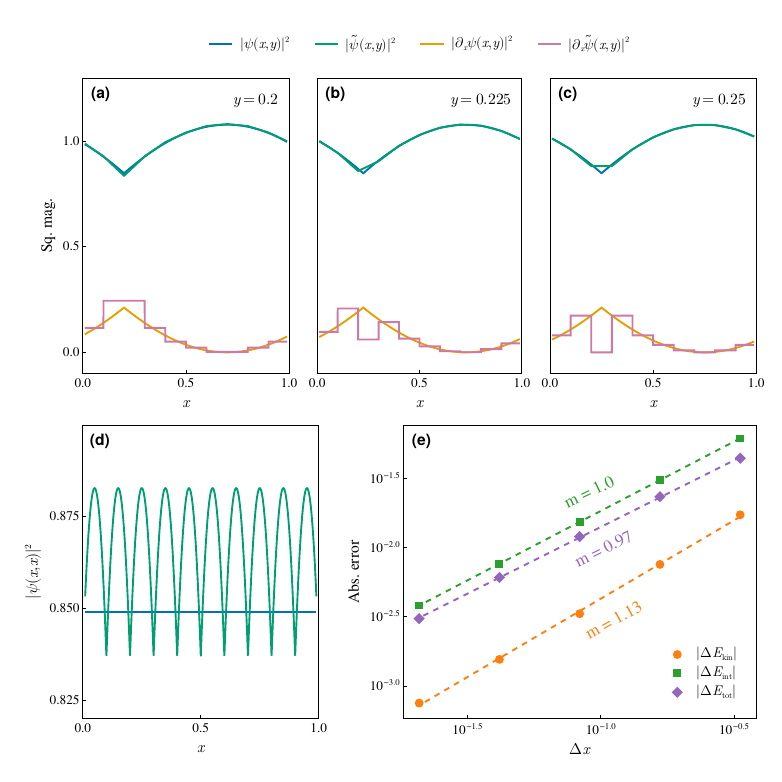}
    \caption{Analysis of the two-particle wavefunction projected onto a periodic tent-function basis. (a--c) Squared magnitudes of the exact and projected wavefunctions $|\psi(x,y)|^2$, alongside their partial derivatives $|\partial_x \psi(x,y)|^2$ appearing in the kinetic energy integral, using $L=10$ tent functions, for fixed positions $y=0.2$, $y=0.225$, and $y=0.25$, which were chosen to either coincide or fall in between the grid points $\{0, 0.1, 0.2, 0.3, \ldots, 1\}$. (d) Squared magnitude $|\psi(x,y)|^2$ evaluated along the diagonal $x=y$, as appearing in the interaction energy integral. (e) Scaling of the absolute errors for the kinetic ($E_{\text{kin}}$), interaction ($E_{\text{int}}$), and total energy ($E_{\text{tot}}$) as a function of the grid spacing $\Delta x = \ell/L$.}
    \label{fig:cusp_error}
\end{figure*}

To explain the linear convergence of the FE-MPS energy as function of grid spacing $\Delta x$, it is instructive to consider the exact Lieb-Liniger ground state with $N=2$. For simplicity, we take a box of length $\ell$ with periodic boundary conditions (PBC). Working in first quantization throughout this section, the Hamiltonian in Eq.~\eqref{eq:LiebLiniger} with $V(x)=0$ for $2$ particles is expressed as
\begin{equation}\label{eq:LiebLinigerFirstQuant}
    H = -\frac{1}{2}\frac{\partial^2}{\partial x_1^2}-\frac{1}{2}\frac{\partial^2}{\partial x_2^2} + 2g \delta(x_1 - x_2).
\end{equation}
The exact wave function is given by \cite{Lieb1963}
\begin{equation}
    \psi(x_1,x_2) = \begin{cases} e^{i(k_1 x_1 + k_2 x_2)}                                  \\
        \quad + S(k_1,k_2) e^{i(k_2 x_1 + k_1 x_2)}, & x_1 < x_2  \\[2ex]
        e^{i(k_1 x_2 + k_2 x_1)}                                  \\
        \quad + S(k_1,k_2) e^{i(k_2 x_2 + k_1 x_1)}, & x_2 < x_1,\end{cases}
\end{equation}
with the scattering matrix defined as
\begin{align}
    S(k_1, k_2) & = \frac{k_1 - k_2 - i \, 2g}{k_1 - k_2 + i \, 2g},
\end{align}
so as to satisfy the cusp condition induced by the $\delta$-interaction \cite{Lieb1963}
\begin{equation}
    \left. \left( \frac{\partial}{\partial x_{2}} - \frac{\partial}{\partial x_1} \right) \psi \right|_{x_{2} = x_1^+} = 2g\psi \Big|_{x_{2} = x_1}.
\end{equation}
The energy of this state is given by
\begin{equation}
    E(k_1,k_2) = \frac{ k_1^2}{2} + \frac{ k_2^2}{2}.
\end{equation}
Imposing PBC as $\psi(x_1+\ell, x_2) = \psi(x_1, x_2)$ and assuming that the ground state has $k_1+k_2=0$, or thus $k_1 = - k_2 = k$, we find $k$ as the solution of
\begin{equation}
    S(k,-k) e^{i k \ell} = 1.
\end{equation}
The lowest energy solution is then found for $k \in [0, \frac{\pi}{\ell}]$, bounded by the limiting values corresponding to $g \to 0^+$ and $g \to +\infty$ respectively.

We can now project $\psi(x_1,x_2)$ onto the tent function basis as
\begin{equation}
    \tilde{\psi}(x_1,x_2) = \sum_{i_1 i_2} C^{i_1 i_2} \phi_{i_1}(x_1) \phi_{i_2}(x_2)
\end{equation}
with
\begin{align}
    C^{i_1 i_2} & = \sum_{j_1j_2}(N^{-1})^{i_1 j_1} (N^{-1})^{i_2 j_2}\nonumber                                                            \\
                & \qquad \times\int_0^\ell \int_0^\ell \overline{\phi_{j_1}(y_1)} \, \overline{\phi_{j_2}(y_2)} \psi(y_1, y_2)\,dy_1 dy_2.
\end{align}

Since this projection amounts to linear interpolation, it can easily be argued and observed that the local error $|\psi(x_1,x_2) - \tilde{\psi}(x_1, x_2)|$ scales as $O(\Delta x^2)$ in the segments where the wave function is smooth, but only scales linearly in $\Delta x$ in the diagonal segments containing $x_1 = x_2$, where the wave function exhibits a cusp. For the derivatives $\partial_{x_1} \psi(x_1,x_2)$ and $\partial_{x_2} \psi(x_1,x_2)$, the order of approximation is one degree lower, meaning in particular that $|\partial_{x_i}\psi(x_1,x_2) - \partial_{x_i} \tilde{\psi}(x_1, x_2)|$ exhibits a $\mathcal{O}(1)$ error in the diagonal segments.

By evaluating the energy contributions of the projected state,
\begin{align}
    \tilde{E}_\text{kin} & = \frac{1}{2} \int_0^\ell \int_0^\ell \left(|\partial_{x_1} \tilde{\psi}(x_1,x_2)|^2 + |\partial_{x_2} \tilde{\psi}(x_1,x_2)|^2\right) \dd x_1 \dd x_2, \\
    \tilde{E}_\text{int} & = g \int_0^\ell  |\tilde{\psi}(x,x)|^2 \dd x,
\end{align}
we find that both terms incur an error relative to the exact solution that scales linearly with $\Delta x$, which is confirmed numerically in Fig.~\ref{fig:cusp_error}. For the kinetic energy, involving a two-dimensional integral, this originates from a local $\mathcal{O}(1)$ error in the integrand in the diagonal segments of area $(\Delta x)^2$, with the number of such segments given by $\ell / \Delta x$. For the interaction energy, the local error in the integrand scales as $O(\Delta x)$ for all $x \in [0, \ell]$. Consequently, the total energy $\tilde{E} = \tilde{E}_{\text{kin}} + \tilde{E}_{\text{int}}$ also converges as $\mathcal{O}(\Delta x)$, demonstrating that the accuracy of the tent basis is constrained by the presence of non-smooth features in the many-body wavefunction.

\section{Observables}
\subsection{Two-point correlators}
These can be evaluated by expanding the field operators in the basis $\{\phi_i(x)\}$ and mapping the expectation value to the computational basis. For example, we consider the first order coherence, $g^{(1)}(x, x')$;
\begin{align}
    \langle \hat{\Psi}^{\dagger}(x) & \hat{\Psi}(x') \rangle = \sum_{i,j} \phi_i^*(x) \langle \Phi | (\hat{b}^i)^{\dagger} \hat{b}^j | \Phi \rangle   \phi_j(x')  \nonumber  \\
                                    & = \sum_{i,j} \phi_i^*(x)  {}_{\ c\!}\langle \Phi | W^{\dagger} (\hat{b}^i)^{\dagger} \hat{b}^j W | \Phi \rangle_c  \phi_j(x')\nonumber \\
                                    & = \sum_{i,j} \phi_i^*(x) {}_{\ c\!}\langle \Phi | c_i^{\dagger} (W^{\dagger} W) c_j | \Phi \rangle_c  \phi_j(x') \nonumber             \\
                                    & = \sum_{i,j} \phi_i^*(x) {}_{\ c\!}\langle \Phi | c_i^{\dagger} \mathcal{N} c_j | \Phi \rangle_c  \phi_j(x')  \nonumber                \\
                                    & = \vec{\phi}^{\dagger}(x) \cdot \mathbb{G}_c \cdot \vec{\phi}(x'),
\end{align}
where $(\mathbb{G}_c)_{ij} = {}_{\,c\!}\langle \Phi | c_i^{\dagger} \mathcal{N} c_j | \Phi \rangle_c$ is the single-particle density matrix (SPDM) calculated in the computational basis, interleaved with the many-body overlap $\mathcal{N}$. Here we assume that ${}_{\,c\!}\langle \Phi |\mathcal{N} | \Phi \rangle_c = 1$. Note that since the tent functions only overlap with their nearest neighbors, all local observables such as the particle density can be computed using just the $i = j$ and $|i - j| = 1$ SPDM elements.

\subsection{Single-particle density matrix}
\label{app:spdm_eval}

Evaluating the SPDM $(\mathbb{G}_c)_{ij} = {}_{\,c\!}\langle \Phi | c_i^{\dagger} \mathcal{N} c_j | \Phi \rangle_c$ requires computing two-point correlations embedded within the background norm MPO $\mathcal{N}$. For a grid with $L$ points, we reduce the computational cost from $O(L^3)$ to $O(L^2)$ by precomputing the standard environments $\mathcal{N}_L^{(k)}$ (Eq.~\eqref{eq:left_env_update}) and $\mathcal{N}_R^{(k)}$ (Eq.~\eqref{eq:right_env_update}) and utilizing a sequential propagation procedure as described below.

The diagonal elements follow from contracting the precomputed environments with the local site operator:
\begin{equation}
    (\mathbb{G}_c)_{ii} =
    \begin{tikzpicture}[baseline=(current bounding box.center), scale=0.55, semithick]
        \node[envtensor] (Lprev) at (0,0.7) {$\mathcal{N}_L^{(i-1)}$};
        \node[mpstensor] (A) at (2.7, 2.2) {$M^{(i)}$};
        \node[mpotensor] (W) at (2.7, 0.7) {$c_i^\dagger\mathcal{N}^{(i)}c_i$};
        \node[mpstensor] (Ab) at (2.7, -0.8) {$\bar{M}^{(i)}$};
        \node[envtensor] (R) at (5.4,0.7) {$\mathcal{N}_R^{(i)}$};
        \draw (Lprev.east |- A) -- (A.west); \draw (Lprev.east |- W) -- (W.west); \draw (Lprev.east |- Ab) -- (Ab.west);
        \draw (A.east) -- (R.west |- A); \draw (W.east) -- (R.west |- W); \draw (Ab.east) -- (R.west |- Ab);
        \draw (A) -- (W) -- (Ab);
    \end{tikzpicture}
\end{equation}
For off-diagonal elements ($i < j$), we proceed row-by-row. For a fixed $i$, we initialize an auxiliary left environment $\tilde{\mathcal{N}}_L^{(i; i)}$ by inserting $c_i^\dagger$ into the background $\mathcal{N}$:
\begin{equation}
    \begin{tikzpicture}[baseline=(current bounding box.center), scale=0.55, semithick]
        \node[altenvtensor] (L) at (0,0.7) {$\tilde{\mathcal{N}}_L^{(i; i)}$};
        \foreach \y in {2.2, 0.7, -0.8} { \draw (L.east |- 0,\y) -- ++(0.7,0); }
    \end{tikzpicture}
    =
    \begin{tikzpicture}[baseline=(current bounding box.center), scale=0.55, thick,
            tensor/.style={draw, rectangle, rounded corners=2pt, minimum width=3em, minimum height=1.8em},
            env/.style={draw, rectangle, rounded corners=2pt, minimum width=3.8em, minimum height=6.5em, fill=gray!5}]
        \node[envtensor] (Lprev) at (0,0.7) {$\mathcal{N}_L^{(i-1)}$};
        \node[mpstensor] (A) at (2.7, 2.2) {$M^{(i)}$};
        \node[mpotensor] (W) at (2.7, 0.7) {$c_i^\dagger \mathcal{N}^{(i)}$};
        \node[mpstensor] (Ab) at (2.7, -0.8) {$\bar{M}^{(i)}$};
        \draw (Lprev.east |- A) -- (A.west); \draw (Lprev.east |- W) -- (W.west); \draw (Lprev.east |- Ab) -- (Ab.west);
        \draw (A.east) -- ++(0.7,0); \draw (W.east) -- ++(0.7,0); \draw (Ab.east) -- ++(0.7,0);
        \draw (A) -- (W) -- (Ab);
    \end{tikzpicture}
\end{equation}
To compute the elements $(\mathbb{G}_c)_{ij}$ for all $j > i$, we propagate this auxiliary environment rightward. At intermediate sites $k$ ($i < k < j$), the environment is updated with the transfer matrix induced by $\mathcal{N}$:
\begin{equation}
    \begin{tikzpicture}[baseline=(current bounding box.center), scale=0.55, semithick]
        \node[altenvtensor] (L) at (0,0.7) {$\tilde{\mathcal{N}}_L^{(i; k)}$};
        \foreach \y in {2.2, 0.7, -0.8} { \draw (L.east |- 0,\y) -- ++(0.7,0); }
    \end{tikzpicture}
    =
    \begin{tikzpicture}[baseline=(current bounding box.center), scale=0.55, thick]
        \node[altenvtensor] (Lprev) at (0,0.7) {$\tilde{\mathcal{N}}_L^{(i; k-1)}$};
        \node[mpstensor] (A) at (2.7, 2.2) {$M^{(k)}$};
        \node[mpotensor] (W) at (2.7, 0.7) {$\mathcal{N}^{(k)}$};
        \node[mpstensor] (Ab) at (2.7, -0.8) {$\bar{M}^{(k)}$};
        \draw (Lprev.east |- A) -- (A.west); \draw (Lprev.east |- W) -- (W.west); \draw (Lprev.east |- Ab) -- (Ab.west);
        \draw (A.east) -- ++(0.7,0); \draw (W.east) -- ++(0.7,0); \draw (Ab.east) -- ++(0.7,0);
        \draw (A) -- (W) -- (Ab);
    \end{tikzpicture}
\end{equation}
The matrix element is finally evaluated at site $j$ by contracting the propagated environment with the annihilation operator and the precomputed right environment:
\begin{equation}
    (\mathbb{G}_c)_{ij} =
    \begin{tikzpicture}[baseline=(current bounding box.center), scale=0.55, semithick]
        \node[altenvtensor] (Lprev) at (0,0.7) {$\tilde{\mathcal{N}}_L^{(i; j-1)}$};
        \node[mpstensor] (A) at (2.7, 2.2) {$M^{(j)}$};
        \node[mpotensor] (W) at (2.7, 0.7) {$\mathcal{N}^{(j)} c_j$};
        \node[mpstensor] (Ab) at (2.7, -0.8) {$\bar{M}^{(j)}$};
        \node[envtensor] (R) at (5.4,0.7) {$\mathcal{N}_R^{(j)}$};
        \draw (Lprev.east |- A) -- (A.west); \draw (Lprev.east |- W) -- (W.west); \draw (Lprev.east |- Ab) -- (Ab.west);
        \draw (A.east) -- (R.west |- A); \draw (W.east) -- (R.west |- W); \draw (Ab.east) -- (R.west |- Ab);
        \draw (A) -- (W) -- (Ab);
    \end{tikzpicture}
\end{equation}
Caching the auxiliary environment during the rightward sweep for each row $i$ ensures $O(L^2)$ complexity. Lower-triangular elements follow from Hermiticity of the SPDM. Similar constructions can be used to compute other two-point observables as well as higher order correlators.

\subsection{Momentum distribution}\label{app:momentum_dist}
The momentum distribution is typically computed as the Fourier transform of the real-space SPDM:
\begin{equation}
    n(k) = \iint \dd x \dd y \, e^{ik(x-y)} \langle \hat{\Psi}^\dagger(x) \hat{\Psi}(y) \rangle.
\end{equation}
Expanding the field operators in the basis $\{\phi_i(x)\}$, we obtain:
\begin{equation}\label{eq:n(k)}
    n(k) = \sum_{j,l} \overline{\tilde\phi_j(k)} (\mathbb{G}_c)_{jl} \tilde\phi_l(k),
\end{equation}
where $\tilde\phi_j(k)$ is the Fourier transform of the tent function $\phi_j(x)$. For the basis defined in Eq.~\eqref{eq:tent_fn}, we have
\begin{equation}
    \tilde\phi_j(k) = \frac{\Delta x^2}{h} e^{-ikx_j} \operatorname{sinc}^2 \left( \frac{k \Delta x}{2} \right).
\end{equation}
Substituting this into Eq.~\eqref{eq:n(k)} yields:
\begin{equation}
    n(k) = \left[ \frac{\Delta x^4}{h^2} \operatorname{sinc}^4 \left( \frac{k \Delta x}{2} \right) \right] \sum_{j,l} (\mathbb{G}_c)_{jl} e^{ik(x_j - x_l)}.
\end{equation}
As $k \to \infty$, the discrete structure factor in the summation remains periodic, while the prefactor provides an envelope decaying as $k^{-4}$. This confirms that the FE-MPS state possesses the correct UV behavior.
In contrast, for FD-MPS, we have
\begin{equation}
    n_{\text{FD}}(k) = \frac{\Delta x}{2\pi} \sum_{j,l} (\mathbb{G}_c)_{jl} e^{ik(x_j - x_l)}.
\end{equation}
This is strictly periodic with period $2\pi/\Delta x$ and cannot recover the expected power-law decay.

\section{Finite difference MPS}\label{app:finite_diff}

In this appendix, we briefly outline the standard procedure for discretizing a continuum field theory, such as the one defined in Eq.~\eqref{eq:LiebLiniger}, onto a lattice for use with MPS techniques. By considering a uniform grid with spacing $\Delta x$ and nodes $x_i$, we can define discretized field operators $\hat{a}_i = \sqrt{\Delta x} \hat{\Psi}(x_i)$ so that $[\hat a_i, \hat a_j^{\dagger}]_{\pm} = \delta_{ij}$. Utilizing a first-order forward difference stencil for the derivative terms, the FD-MPS Hamiltonian for Eq.~\eqref{eq:LiebLiniger} takes the form:
\begin{align}
    \hat{H}_{\text{FD}} & = \sum_i \left[ \left( \frac{1}{(\Delta x)^2} + V(x_i) - \mu \right) \hat{a}_i^\dagger \hat{a}_i \right. \nonumber                                                                         \\
                        & \quad \left. - \frac{1}{2 (\Delta x)^2} \left( \hat{a}_i^\dagger \hat{a}_{i+1} + \text{h.c.} \right) + \frac{g}{\Delta x} \hat{a}_i^\dagger \hat{a}_i^\dagger \hat{a}_i \hat{a}_i \right],
\end{align}
where $\hbar=m=1$. Since we use a first-order stencil, the error in energy is expected to vanish as $\mathcal{O}(\Delta x^2)$. While this treatment is sufficient for fermionic systems, the application to bosonic operators requires truncating the infinite local Hilbert space at a particle number cutoff $n_c$ per site. However, this naive truncation introduces a subtle numerical artifact due to an interplay between the cutoff $n_c$ and the grid spacing $\Delta x$ that generates significant kinetic energy errors scaling as $\mathcal{O}(\Delta x^{n_c/2+1})$ \cite{ganahl_continuous_2018}. These errors arise from an unbalanced cancellation of terms at the edge of the local Hilbert space, which can be mitigated by modifying the kinetic energy through a local projector $\hat{P}_i = \mathbb{I} - \ket{n_c}\bra{n_c}$ that explicitly removes the highest occupancy state from the derivative.

By replacing the standard first-order derivative $\hat{\partial}_x \Psi(x_i) \approx (\Delta x)^{-3/2}(\mathbb{I}_i \otimes \hat{a}_{i+1} - \hat{a}_{i} \otimes \mathbb{I}_{i+1})$ with the projected form $(\Delta x)^{-3/2}(\hat{P}_i \otimes \hat{a}_{i+1} - \hat{a}_{i} \otimes \hat{P}_{i+1})$, the resulting Hamiltonian is expressed as:
\begin{align}
    \hat{H}_{\text{FD}} & = \sum_i \left[ \left(V(x_i) - \mu \right) \hat{a}_i^\dagger \hat{a}_i\right. \nonumber                                                                                                    \\
                        & \quad + \frac{1}{2(\Delta x)^2}(\hat{a}_i^\dagger \hat{a}_i \otimes \hat{P}_{i+1} + \hat{P}_{i} \otimes \hat{a}_{i+1}^\dagger \hat{a}_{i+1}) \nonumber                                     \\
                        & \quad \left. - \frac{1}{2 (\Delta x)^2} \left( \hat{a}_i^\dagger \hat{a}_{i+1} + \text{h.c.} \right) + \frac{g}{\Delta x} \hat{a}_i^\dagger \hat{a}_i^\dagger \hat{a}_i \hat{a}_i \right].
\end{align}
This formulation remains faithful to the continuum physics while eliminating the superfluous errors inherent in the truncation scheme, allowing for high-precision results with a significantly smaller local dimension (e.g., $n_c < 5$) than the $n_c \ge 10$ typically required in previous studies.

Furthermore, as discussed in Ref.~\cite{ganahl_continuous_2018}, one can pursue an even more aggressive reduction of the local Hilbert space by splitting the interaction term into a nearest-neighbor form. By replacing the on-site interaction with a term of the form $\frac{g}{\Delta x} \hat{a}_i^\dagger \hat{a}_i \hat{a}_{i+1}^\dagger \hat{a}_{i+1}$, the continuum limit is preserved as $\Delta x \to 0$. This modification allows the local Hilbert space to be reduced to $n_c=1$, effectively representing the system as hard-core bosons on a lattice. However, this modification comes at the cost of degrading the asymptotic convergence of energy from $\mathcal{O}(\Delta x^2)$ to $\mathcal{O}(\Delta x)$.
\section{Bethe ansatz}\label{app:bethe_ansatz}

To benchmark FE-MPS against semi-analytical results, we employ the coordinate Bethe ansatz for solving the Lieb-Liniger model \cite{zvonarev2010}. This is conventionally formulated for the Hamiltonian $H_{LL} = -\sum \partial_x^2 + 2c\sum_{i<j} \delta(x_i - x_j)$ \cite{Lieb1963}; however, we instead utilize Eq.~\eqref{eq:LiebLinigerFirstQuant} to preserve our choice of units ($\hbar = m = 1$). The numerical implementations used to generate the reference data are available on Github \cite{betheansatz_pkg}. We briefly outline the details as follows.

\subsection{Infinite well}
For a finite system of $N$ bosons in a one-dimensional box of length $L$, the ground state rapidities $\{k_j\}$ satisfy the discrete Bethe equations \cite{Batchelor_2005}:
\begin{equation}
    2Lk_j = 2\pi n_j - \sum_{l \neq j}^N \left[ \theta(k_j - k_l) + \theta(k_j + k_l) \right],
\end{equation}
where $n_j \in \{1, 2, \dots, N\}$ and the scattering phase shift is given by
\begin{equation}
    \theta(x) = 2 \arctan\left(\frac{x}{2g}\right).
\end{equation}
Since the corresponding Jacobian is symmetric positive-definite for $g>0$, we can efficiently solve this non-linear system with a few Newton-Raphson iterations. The ground state energy is then computed as $E = \frac{1}{2}\sum_j k_j^2$.

\subsection{Non-uniform potentials}
For non-uniform potentials, such as the harmonic trap, we utilize the Local Density Approximation (LDA) combined with the thermodynamic limit (TDL) of the Bethe equations. Assuming local homogeneity, the local chemical potential is given by $\mu(x) = \mu_0 - V(x)$.
For a given $\mu(x)$, the local rapidity density $\rho(k)$ on the interval $[-Q, Q]$ is determined by the integral equation \cite{zvonarev2010}:
\begin{equation}\label{eq:tdl_ba}
    \rho(k) - \frac{1}{2\pi} \int_{-Q}^{Q} K(k - q) \rho(q) \, dq = \frac{1}{2\pi},
\end{equation}
where the kernel $K(x) = d\theta/dx = 4g/(4g^2 + x^2)$. The local particle density $n(x)$ and internal energy density $e(x)$ are then computed as
\begin{equation}
    n = \int_{-Q}^Q \rho(k) \, dk, \quad e = \frac{1}{2}\int_{-Q}^Q k^2 \rho(k) \, dk.
\end{equation}

Eq.~\eqref{eq:tdl_ba} is a Fredholm equation of the second kind and can be solved using standard numerical methods \cite{Fredholm2019}. Finally, the global chemical potential $\mu_0$ is tuned iteratively to satisfy the total particle number constraint $N_{tot} = \int n(x) \, dx$. Once $\mu_0$ is found, the total ground state energy is evaluated as $E_{total} = \int \left[ e(x) + V(x)n(x) \right] dx$. In order to avoid excessive computational cost, we utilize a Chebyshev interpolant as a surrogate for the true $e(x)$ and $n(x)$ when computing the integrals.

\section{Alternate approaches to work with non-orthogonal basis}
\label{app:other_approaches}

When utilizing a non-orthogonal basis to solve many-body ground state problems, several strategies can be employed. Each involves a fundamental trade-off between Hamiltonian complexity and the structure of the underlying Hilbert space. We outline some alternatives to our approach in this Appendix.

\subsection{Orthonormal basis}

The most common approach is to transform the non-orthogonal set $\{\phi_i\}$ into an orthonormal basis $\{\chi_i\}$ through a linear transformation of the operators, $\vec a^{\dagger} = \vec c^{\dagger} F$, as described in Eq.~\eqref{eq:defW}. Under this basis change, the Hamiltonian matrix elements derived in \appref{app:hamiltonian_specifics} transform as:
\begin{align}
    \tilde{t}_{ij}   & = (F^{-\dagger})_i^{\,k} \, t_{kl} \, (F^{-1})^l_{\,j},                                                 \\
    \tilde{U}_{ijkl} & = (F^{-\dagger})_i^{\,p} \, (F^{-\dagger})_j^{\,q} \, U_{pqrs} \, (F^{-1})^r_{\,k} \, (F^{-1})^s_{\,l}.
\end{align}
In the main text, we note that a single-particle overlap $N$ with bandwidth $R$ results in an upper triangular $F$ with the same bandwidth using the Cholesky decomposition. While $F$ is local, its inverse $F^{-1}$ is dense, albeit with entries that decay exponentially away from the diagonal. Consequently, the resulting Hamiltonian in the orthonormal basis is no longer strictly local. In such cases, one must still impose an explicit truncation on the magnitude of the Hamiltonian terms to keep the MPO bond dimension manageable. On the other hand, our approach maintains locality by construction, and the resulting bond dimension does not scale with the system size $L$.

\subsection{Biorthonormal basis}
Alternatively, one may formulate the Hamiltonian in a mixed form using the original creation operators $\hat{a}_i^{\dagger}$ and the dual annihilation operators $\hat{b}^j$. Since these operators satisfy $[\hat{b}^i, \hat{a}_j^{\dagger}] = \delta^i_j$, they can be treated numerically as standard operators that satisfy the CCR. The Hamiltonian is then expressed as:
\begin{equation}
    \hat{H} = \sum_{ij} \tilde{t}^i_j \hat{a}^{\dagger}_i \hat{b}^j + \sum_{ijkl} \tilde{U}^{ij}_{kl} \hat{a}^{\dagger}_i \hat{a}^{\dagger}_j \hat{b}^l \hat{b}^k,
\end{equation}
where the coefficients transform as:
\begin{align}
    \tilde{t}^i_j       & = \sum_k (N^{-1})^{ik} t_{kj},                     \\
    \tilde{U}^{ij}_{kl} & = \sum_{m,n} (N^{-1})^{im} (N^{-1})^{jn} U_{mnkl}.
\end{align}
As this approach utilizes different representations for the bra and ket components, the resulting second-quantized Hamiltonian is no longer Hermitian ($\hat{H} \neq \hat{H}^{\dagger}$). Consequently, the DMRG algorithm must be generalized to solve for distinct right and left eigenvectors:
\begin{align}
    \hat{H} \ket{\psi_R} = E \ket{\psi_R}, \quad \hat{H}^{\dagger} \ket{\psi_L} = \bar{E} \ket{\psi_L}.
\end{align}
Observables are then computed using the biorthogonal expectation value $\langle \hat O \rangle = \langle \psi_L | \hat{O} | \psi_R \rangle / \langle \psi_L | \psi_R \rangle$. This method also introduces non-local couplings due to the insertion of $N^{-1}$. Additionally, since the Hamiltonian is non-Hermitian, we also lose variationality.

\bigskip \noindent
We see that both formulations seek to restore a standard Hilbert space metric by embedding the non-orthogonality into the single-particle operators, effectively trading locality for a simplified Hilbert space structure. This stands in contrast to the approach that we use in this work, where we maximally preserve the physical locality of the Hamiltonian and account for the small degree of non-locality induced by the finite element basis through the non-trivial many-body overlap.

\bibliography{TentMPS.bib}
\end{document}